\documentclass[10pt,aps,pre,noshowpacs,twocolumn,superscriptaddress,nobibnotes,nofootinbib]{revtex4} 

\usepackage{amssymb}
\usepackage{graphicx}
\usepackage{amsmath,amsfonts}
\usepackage[colorlinks=true,allcolors=blue]{hyperref}
\usepackage[utf8]{inputenc}
\usepackage{xcolor}  
\usepackage[normalem]{ulem}
\usepackage[T1]{fontenc} 
\usepackage{multirow}
\usepackage{booktabs}
\usepackage{threeparttable}
\usepackage{tabularx}

\usepackage{wasysym}

\usepackage{pdfpages}

\begin{document}

\title{Chordless cycle filtrations for dimensionality detection in complex networks via topological data analysis}
\date{\today}

\author{Aina Ferr\`a Marc\'us}
\affiliation{Departament de Mat\`ematiques i Inform\`atica, Universitat de Barcelona,Gran Via de les Corts Catalanes 585, 08007 Barcelona, Spain}
\author{Robert Jankowski}
\affiliation{Departament de F\'isica de la Mat\`eria Condensada, Universitat de Barcelona, Mart\'i i Franqu\`es 1, E-08028 Barcelona, Spain}
\affiliation{Universitat de Barcelona Institute of Complex Systems (UBICS), Universitat de Barcelona, Barcelona, Spain}
\affiliation{Faculty of Electrical Engineering, Mathematics and Computer Science, Delft University of Technology, 2628 CD, Delft, Netherlands}
\author{Meritxell Vila Miñana}
\affiliation{Center for Complex Networks and Systems Research, Luddy School of Informatics, Computing, and Engineering, Indiana University, Bloomington, IN, USA}
\author{Carles Casacuberta}
\affiliation{Departament de Mat\`ematiques i Inform\`atica, Universitat de Barcelona,Gran Via de les Corts Catalanes 585, 08007 Barcelona, Spain}
\author{M. {\'A}ngeles Serrano}
\affiliation{Departament de F\'isica de la Mat\`eria Condensada, Universitat de Barcelona, Mart\'i i Franqu\`es 1, E-08028 Barcelona, Spain}
\affiliation{Universitat de Barcelona Institute of Complex Systems (UBICS), Universitat de Barcelona, Barcelona, Spain}
\affiliation{ICREA, Passeig Llu\'is Companys 23, E-08010 Barcelona, Spain}
\email[]{marian.serrano@ub.edu}

\begin{abstract}
Many complex networks, ranging from social to biological systems, exhibit structural patterns consistent with an underlying hyperbolic geometry.
Revealing the dimensionality of this latent space can disentangle the structural complexity of communities, impact efficient network navigation, and fundamentally shape connectivity and system behavior. We introduce a topological data analysis weighting scheme for graphs based on chordless cycles to estimate network dimensionality in a data-driven way. We further show that the resulting descriptors can effectively estimate network dimensionality using a neural network architecture trained on a synthetic graph database constructed for this purpose, which requires no retraining to transfer effectively to real-world networks. Thus, by combining cycle‐aware filtrations, algebraic topology, and machine learning, our approach provides a robust and effective method for uncovering the hidden geometry of complex networks and guiding accurate modeling and low-dimensional embedding.
\end{abstract}

\maketitle
\let\oldaddcontentsline\addcontentsline
\renewcommand{\addcontentsline}[3]{}

\section{Introduction}
\label{introduction}
The rapid growth of data has created significant challenges in science and technology. Large datasets from fields such as biology (e.g., gene expression and protein interactions), social networks (e.g., agent behavior), and physics (e.g., cosmological simulations) often contain rich structural information hidden in their complex nature. Capturing this information requires tools capable of detecting patterns in data. Topological Data Analysis (TDA)~\cite{ballester2024on,carlsson2021topological,edelsbrunner2022,wasserman2018topological} is one such approach that uses ideas from topology to identify features that persist across scales.

In TDA, scales are usually defined through filtrations, a systematic way to build a sequence of simplicial complexes that encode the geometric and topological structure of a dataset across multiple levels of resolution. Central to the analysis of such simplicial complexes is persistent homology~\cite{carlsson2021topological,edelsbrunner2022}, a main TDA computational method that tracks the emergence, persistence, and disappearance of topological features ---such as connected components or loops---
in a filtered simplicial complex as the filtration parameter evolves. In this context, graphs can be treated as simplicial complexes and filtered by assigning weights to their nodes and/or edges.

The success of persistent homology depends critically on the choice of a filtration and topological features of interest, and the selection of these depends on the problem being addressed. Not every filtration or feature descriptor is suitable for every problem, underscoring the importance of designing suitable TDA pipelines to achieve meaningful results. 
The triad filtration-feature-framework (FFF) embodies an intertwined relationship that forms the foundation for effectively leveraging persistent homology. However, filtration schemes based on a variety of quantifiers, such as Forman--Ricci curvature~\cite{forman2003bochner} or betweenness centrality~\cite{freeman1977set}, are sometimes applied indiscriminately to data without considering the specifics of the selected topological features and the nature or framework of the problem at hand. This work emphasizes the importance of aligning the FFF triad by addressing dimensionality detection in complex networks, thereby bridging the gap between complex network theory and TDA.
Simplicial complexes associated with complex networks have emerged as powerful representations in network science \cite{bianconi2021higher,bobrowski2022random,hernandez2020simplicial,salnikov2018simplicial,torres2020simplicial}, offering insights into higher-order structures beyond pairwise interactions. Moreover, persistent homology of filtered simplicial complexes has been applied to the study of complex networks in general~\cite{horak2009,taylor2015topological,kannan2019,Myers2019,guerra2021,jhun2022} and to neuroscience in particular~\cite{petri2014,giusti2015,sizemore2019importance,das2023}.

The problem addressed, the detection of the effective dimensionality of the space in which a complex phenomenon unfolds, is a recurring theme across the sciences. In statistical physics, the spatial dimension strongly constrains scaling laws and often determines the universality class of critical, extended systems in which multiple length scales are relevant. Even in the Ising model ---the simplest widely used model of collective behavior and phase transitions, in which binary spins tend to align with their neighbors---
dimensionality has a strong impact on critical behavior~\cite{ising1925,onsager1944}. 
Diffusion processes are also often studied through random walks on $D$-dimensional lattices, where the dimension controls return probabilities, exploration rates, and typical first-passage times~\cite{redner2001,dvoretzky1951,condamin2007}. In other areas of physics, such as string-theory settings, extra dimensions beyond the observed three are needed, but remain inaccessible because they are tightly compactified or because observable dynamics are confined to a lower-dimensional subset~\cite{polchinski1998}. A similar idea appears in computer science: although data and interactions may live in a formally high-dimensional space, relevant configurations typically occupy a much smaller effective subspace, with important implications for both structure and dynamics~\cite{tenenbaum2000,belkin2003}. Dimensionality is also crucial in biology: transport and diffusion of molecules, encounter rates, and search processes depend sensitively on effective dimension~\cite{codling2008,vonhippel1989}, and the structure of regulatory, neuronal, and ecological networks can enforce low-dimensional or genuinely multiscale dynamics, with different dimensions governing local versus global behavior~\cite{peach2022}. Thus, the effective dimensionality of complex phenomena has direct implications for structural properties, dynamical regimes, and the mechanisms that dominate robustness, controllability, and criticality in complex systems.

In network science, several non-equivalent definitions of dimensionality that probe different aspects of structure and dynamics have been proposed in the literature. Topological shortest pahts have been used to define chemical or fractal (box-counting)  dimension~\cite{eguiluz2003effective,song2007calculate,shanker2007defining,wei2014new,kim2007box}, including networks with explicit geometrical embedding~\cite{daqing2011dimension}; related to this, metric dimension is defined as the smallest number of nodes required to identify all other nodes uniquely based on shortest path distances~\cite{tillquist2023getting}; correlation dimension is computed from network trajectories of random walkers~\cite{lacasa2013correlation}; and spectral dimension depends on the spectral properties of the graph Laplacian~\cite{PhysRevLett.76.1091,avrachenkov2019eigenvalues}. The problem has also been addressed within the framework of network geometry~\cite{boguna2021network}. A~model-driven approach~\cite{almagro2022detecting} leverages the geometric $\mathbb{S}^D/\mathbb{H}^{D+1}$ model~\cite{serrano2008self,budel2024random}, which reproduces the observed connectivity of real networks, to reveal their intrinsic dimensionality in a latent hyperbolic space, where nodes are more likely to be connected if they are closer to each other.  The real network's specific dimension is determined by projecting the frequencies of chordless cycles of varying lengths onto the background model's statistical configuration space. This technique revealed ultra-low-dimensional structures in real networks previously masked by apparent high-dimensionality~\cite{almagro2022detecting}. This includes, for instance, tissue-specific biomolecular networks being extremely low-dimensional, brain connectomes being close to the three dimensions of their anatomical embedding, and social networks and the Internet, requiring slightly higher dimensionality. Within the same modeling framework, an alternative method for determining network dimensionality uses an embedding technique, named $D$-Mercator~\cite{jankowski2023d}, to produce multidimensional maps of real networks in $(D+1)$-hyperbolic space whose dimensionality strongly shapes network properties, including community diversity~\cite{budel2024random,desy2023}. The maps are used to estimate intrinsic dimensionality based on navigability and community structure, yielding results consistent with the statistical configuration space approach.

Despite this broad relevance, the concept of network dimensionality remains largely unexplored within TDA. In this work, we propose a method for dimensionality detection in complex networks based on TDA, which complements a trilogy in combination with the configuration-space and embedding methods used in~\cite{almagro2022detecting} and~\cite{jankowski2023d}, respectively. 
The dimensionality detection task, crucial for understanding the intrinsic geometry of data, showcases how tailored filtrations and feature selection can enhance the ability of persistent homology to analyze and interpret complex data for specific purposes. 
Specifically, we introduce a chordless cycle filtration scheme and use it to compute extended persistence of cycles, as the topological descriptor that best captures the distribution of cycles in synthetic and real networks to predict their dimensionality. 

Moreover, we propose a data-driven approach to estimate network dimensionality by training a neural network on nearly $800\,000$ synthetic networks. A~multilayer perceptron accurately estimates dimensionality and transfers effectively to real-world networks, generalizing and adapting to new data without retraining. Ablation experiments demonstrate that TDA features play an important role in this task, even when combined with average cycle densities and degree-related graph features.

\section{Results}
Our approach to estimating network dimensionality using persistent homology is twofold. In \cite{almagro2022detecting}, it has been shown that measuring the intrinsic dimensionality of a complex network is possible by computing profiles of structural properties that are sensitive to dimensionality. These properties are densities of chordless cycles of sizes three (triangle), four (squares), and five (pentagons), but the method required generating ensembles of synthetic networks for each candidate network. Here, we instead focus on persistence summaries using a filtration based on the density of cycles and show that these descriptors reliably detect latent dimensionality. We then leverage the descriptors to develop a supervised machine‐learning model, trained on a large database of synthetic graphs with known dimensions and controlled properties, to predict the intrinsic dimension of an input network. 

\subsection{Persistence descriptors for dimensionality estimation}
\label{subsec:data-driven}
Many real networks share universal properties, such as sparsity, heavy-tailed degree distributions, the small-world effect, high clustering coefficients, and self-similarity. These properties can be captured by a simple geometric framework~\cite{serrano2021shortest} using the $\mathbb{S}^1/\mathbb{H}^2$ model~\cite{serrano2008self,krioukov2010hyperbolic}, which combines a popularity coordinate ---controlling node degrees--- with a similarity coordinate that represents all other attributes influencing network connectivity. This model can be generalized to a $D$-dimensional similarity space, yielding the $\mathbb{S}^D/\mathbb{H}^{D+1}$ model~\cite{serrano2008self,budel2024random}; see Section~\ref{sec:methods} for more details. These models have been used to determine the dimensionality of real networks by analyzing their cycle profiles~\cite{almagro2022detecting}. Here, however, we propose to measure their persistence descriptors instead.

The goal of persistent homology is to compute topological features of a space equipped with a filtration. In the context of graphs, one can define filtrations by endowing either nodes or edges with suitable weightings. A common choice is the degree filtration on nodes; however, it has been shown that the degree filtration is less expressive for some graph learning tasks than other, motif-based, filtrations~\cite{ballester2024on}. In this work, we propose an edge weighting scheme based on densities of chordless cycles. A~chordless cycle is defined as a closed edge path in which no two non-consecutive nodes are connected by an edge.

Our goal is to determine an optimal value of $D$ for the $\mathbb{S}^D/\mathbb{H}^{D+1}$ model corresponding to a real network, in which $\gamma$ and $\beta$ ---the parameters of the model controlling the scale-free degree distribution and the clustering coefficient, respectively--- are usually unknown. In our data-driven method, optimality is defined in the same way as in~\cite{almagro2022detecting}, by picking the dimension of a closest point in the configuration space of synthetic surrogates generated from the given real network, using $D$-dimensional geometric randomization (D-GR), as described in Section~\ref{subsec:hyperbolic_models}, with different values of $\gamma$, $\beta$, and~$D$. The D-GR model, originally proposed for $D=1$~\cite{starnini2019geometric}, works on the observed sequence of node degrees and rewires the network to maximize the likelihood that the new topology is generated by the $\mathbb{S}^D/\mathbb{H}^{D+1}$ model.

Thus, for each complex network $G_0$, whose dimensionality is to be estimated, we generated an ensemble of synthetic surrogates $G_1,\dots,G_n$ using the $\mathbb{S}^D$ model with a range of values of $D$ and a range of values of the clustering coefficient $\beta$ until $\beta=6D$. 
Densities of edge triangles, chordless squares, and chordless pentagons were then computed as specified in Section~\ref{subsec:features} for each surrogate in the ensemble and for the target network.
Therefore, each network $G$ yields three weighted graphs $(G,w_t)$, $(G,w_s)$ and $(G,w_p)$, where the edge weights $w_t$, $w_s$, $w_p$ are the densities of triangles, squares, and pentagons, respectively. The mean value of $w_t$ over all the edges of a graph is denoted by~$C_t$, and similarly with squares and pentagons.

\begin{figure}[h!]
    \centering
    \includegraphics[width=0.45\textwidth]{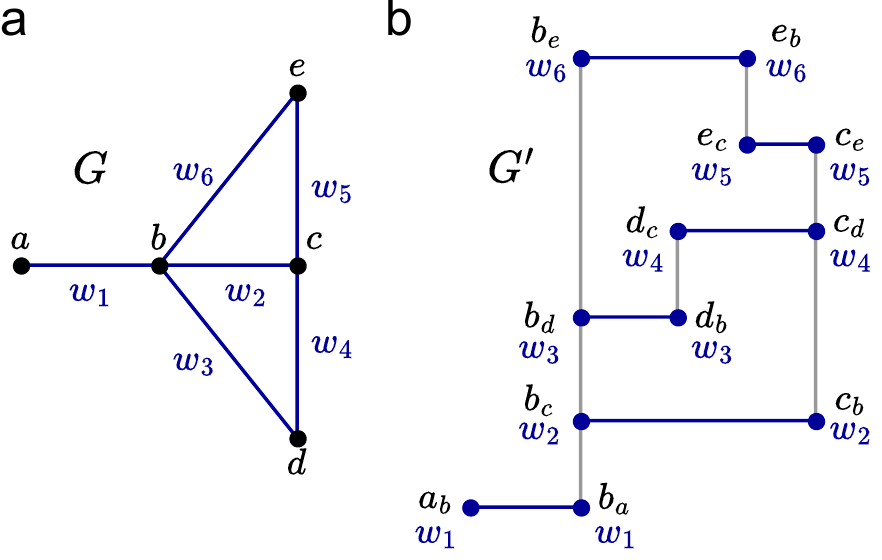}
    \caption{\textbf{Extended persistence for an edge-weighted graph filtration.}
        Topologically equivalent graphs equipped with (a) an edge weighting and (b) a node weighting, with the same extended persistence barcode. Assuming that $w_1<\dots<w_6$, there are two generating cycles with birth-death coordinates $(w_4,w_2)$ and $(w_6,w_3)$, yielding a total extended persistence of $|w_2-w_4|+|w_3-w_6|$. See Section~\ref{subsec:features} for details. }
    \label{fig:blow-up} 
\end{figure}   

Topological features of graphs equipped with a filtration given by edge weights were computed using persistent homology, a tool from algebraic topology used to describe shape characteristics from many kinds of data (Section~\ref{subsec:features}). In this work, persistence refers specifically to the evolution of cycles along the values of a given filtration.
However, cycles in graphs have infinite persistence, since there are no higher-dimensional simplices to eventually fill them. We therefore computed extended persistence of cycles, defined as the difference between the largest and smallest weights among the simplices forming a closed path (see Fig.~\ref{fig:blow-up}). Since we focus on graphs equipped with edge weightings only, we introduce in this article a technique for replacing a given edge-weighted graph with a larger, topologically equivalent graph, 
that carries weights on nodes and on the original edges in a way that is consistent with both sublevel and superlevel filtrations. This is illustrated in Fig.~\ref{fig:blow-up} and explained with more detail in Section~\ref{subsec:features}. 

\begin{figure}[ht!]
    \centering
    \includegraphics[width=0.45\textwidth]{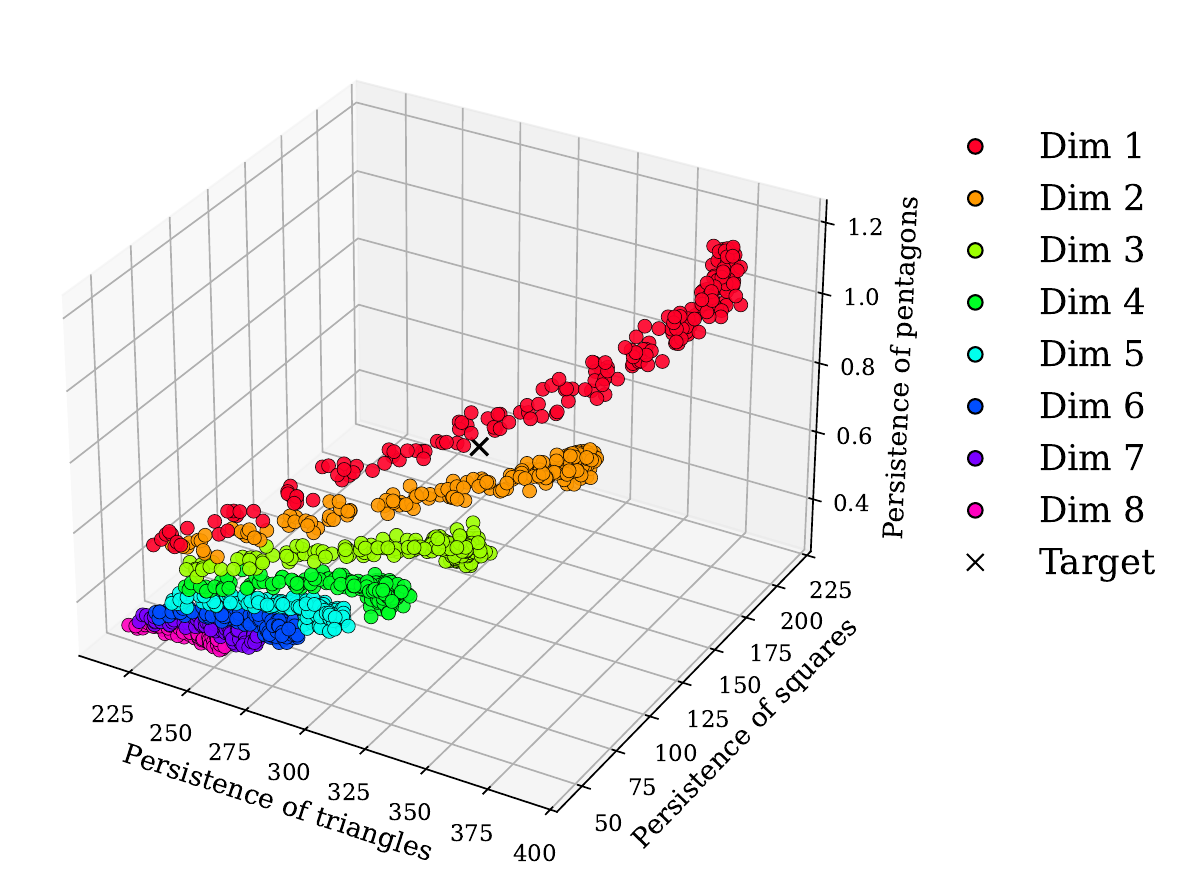}
    \caption{\textbf{Total persistence values obtained from the human connectome.} 3D view of a point cloud representing an ensemble of $1330$ surrogates of the Human2 network (connectome of the human brain, including one hemisphere) in the configuration space of total persistence computed from three chordless cycle densities (triangles, squares, and pentagons). Points are coloured by dimension. The target network is marked with a black cross.} 
    \label{fig:Human2-3D}
\end{figure}

Total extended persistence was used as a topological descriptor, resulting in a feature vector $(TP_t, TP_s, TP_p)$ for each network. A~representation of the configuration space is shown in Fig.~\ref{fig:Human2-3D}, where each point corresponds to a surrogate graph. Points are coloured by the corresponding dimension. The target network from which surrogates were generated has $D=1$ and is marked with a black cross. 

Plausibility of an association between total extended persistence of cycles in a graph and its embedding dimensionality is illustrated in Figs.~\textcolor{blue}{S1}, \textcolor{blue}{S2}, and \textcolor{blue}{S3}, which show that dimensionality is inversely associated with total extended persistence values. A~negative correlation is more evident in the case of squares and pentagons, with respective Pearson coefficient values of $\rho_s=-0.3140$ and $\rho_p=-0.2433$, versus $\rho_t=-0.0522$ for triangles, calculated using a sample of size $50\,000$ from the \textsc{SynNet} database (Section~\ref{subsec:neural_networks}).

\begin{figure*}[ht!]
    \centering
    \includegraphics[width=0.8\textwidth]{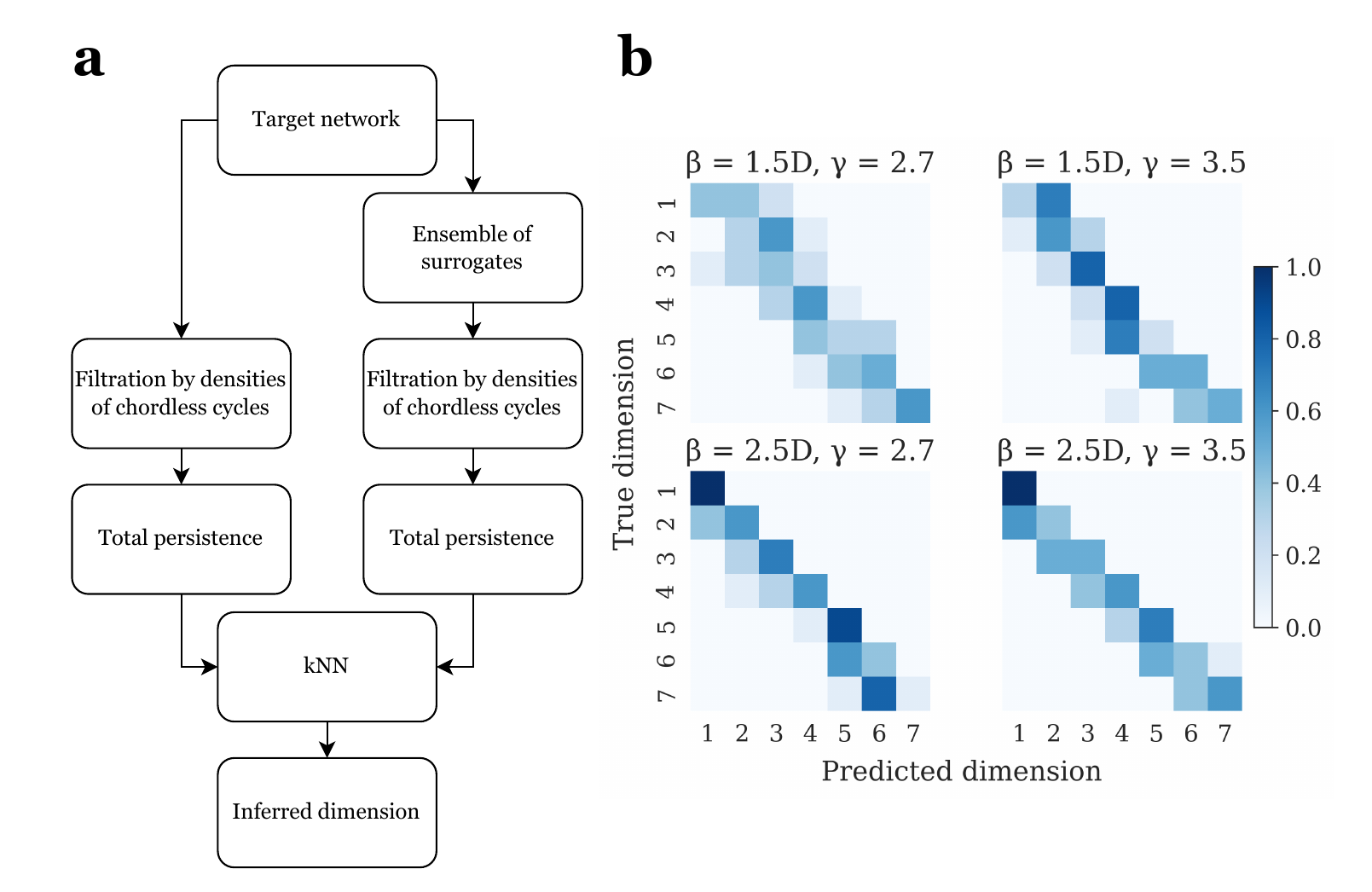}
    \caption{\textbf{Dimensionality estimation with persistence descriptors.} (a) Pipeline of our first method and (b) confusion matrices. Given a target network, we generate an ensemble of surrogates and, for each of them (including the given network), we compute densities of chordless cycles. We equip the graphs with these weightings and perform a topological analysis using extended persistence. For each network, we compute the total persistence of cycles and use a kNN classifier in the configuration space to infer a dimensionality $D^*$ for the real network. The upper two confusion matrices show results obtained with synthetic networks for $\beta = 1.5\,D$ and two different values $\gamma = 2.7$ and $\gamma = 3.5$. This choice of $\beta$ corresponds to the small-world phase even if $\gamma > 3$. The lower two confusion matrices show results for $\beta = 2.5\,D$; in this case, networks with $\gamma > 3$ are large worlds. In each matrix, the rows correspond to the true dimension $D$, and the columns correspond to the inferred dimension~$D^*$. Thus, the $j$-th box in the $i$-th row shows the fraction of $i$-dimensional networks that were classified as $j$-dimensional. The darker the color, the greater the number. Each matrix is evaluated with 70 synthetic networks.}
    \label{fig:pipeline} 
\end{figure*}

In order to estimate the dimensionality of the target network, a $k$-nearest neighbors classifier (kNN) was used. The classifier identifies the $k$ surrogates closest to the target network in the surrogate configuration space $(TP_t, TP_s, TP_p)$ by minimizing Euclidean distance. 
The value of $k$ was not fixed, but it was determined within each ensemble of surrogates, by finding the value of $k$ with highest accuracy in classifying each surrogate in the ensemble. Specifically, the inferred dimension $D^*$ of the target network $G_0$ maximizes the weighted frequency $ f(D) = \sum_{i=1}^{k} \omega_i \,\delta_{D_i, D},$ where the normalized weights are inversely proportional to the distance between the real network and the $i$-th surrogate $G_i$ in the $(TP_t, TP_s, TP_p)$ space, and $\delta_{D_i, D}$ is the Kronecker delta function \cite{almagro2022detecting}. A schematic pipeline summary of the suggested methodology is shown in Fig.~\ref{fig:pipeline}a. 

The closeness of the kNN approximation was measured with a congruency index reflecting the fidelity of the surrogates with respect to the target network. The index is defined as the ratio $\rho=d_0/\overline{d}$, where $d_0$ is the distance between the point in configuration space corresponding to the target network and its closest neighbor, and $\overline{d}$ is the average of the distances between each surrogate point and its closest neighbor.
Hence, the value of $\rho$ is large when the point cloud is clustered while the target network falls far away from the clusters. For comparability reasons, it is convenient to provide values of $1/\rho$, as in Table~\ref{tab:summary_dimension}, since, in most cases, $1/\rho$ takes values between $0$ and~$1$, with values closer to one indicating higher fidelity between the target network and its surrogates.

\begin{table*}[ht]
    \centering
    \caption{\textbf{Inferred dimensionality of real-world networks.} Comparison of the inferred dimension between three different methods: (1)~Mean densities of chordless cycles; (2)~TDA: Persistent homology from chordless cycle density filtrations; (3)~\textsc{DimNN}: A~neural network trained on a database of synthetic networks (introduced in Section~\ref{subsec:neural_networks}).
        The $1/\rho$ values are inverses of the congruency indices; higher values correspond to a closer match between the real network and its surrogate models.}
    \begin{tabular}{@{} llccccc @{}}
        \toprule
        \multicolumn{1}{@{}l}{\textbf{Network}} 
        & \textbf{Domain} 
        & \multicolumn{2}{c}{\textbf{Densities}} & \multicolumn{2}{c}{\textbf{TDA}} & \textbf{\textsc{DimNN}}\hspace{0.4cm}\\
        \cmidrule(lr){3-7}
        &  & \textbf{dim} & $1/\rho$ & \textbf{dim} & $1/\rho$ & \hspace{0.2cm}\textbf{dim}\hspace{0.2cm} \\
        \midrule
        Human2-C    & Biological (Connectome)  & 3 & 0.07 & 1 & 0.21 & 2 \\
        Human-M     & Biological (Metabolic)   & 3 & 0.19 & 4 & 0.46 & 3 \\
        Cargoships  & Economic (Trade)         & 3 & 0.10 & 2 & 0.09 & 5 \\
        Bible-CO    & Informational (Language) & 4 & 0.05 & 1 & 0.04 & 5 \\
        Jazz-CA     & Social (Collaboration)   & 2 & 0.06 & 2 & 0.41 & 2 \\
        EUEmail     & Social (Communication)   & 2 & 0.09 & 1 & 0.28 & 2 \\
        Friends-ON  & Social (Communication)   & 6 & 0.06 & 6 & 0.19 & 7 \\
        Friends-OFF & Social (Offline)         & 8 & 0.72 & 8 & 0.28 & \hspace{-0.2cm}10 \\
        \bottomrule
    \end{tabular}
    \label{tab:summary_dimension}
\end{table*}

As an evaluation step, the performance of the chordless cycle density filtration was tested on synthetic target networks generated using the $\mathbb{S}^D$ model for specific values of $D$ (from $1$ to~$7$), with different degree heterogeneities by varying the exponent ($\gamma = 2.7$ and $\gamma = 3.5$), and  two values of inverse temperature ($\beta = 2.5\,D$, corresponding to the high clustering regime, and $\beta = 1.5\,D$, corresponding to the low clustering regime).
For each combination $(D,\gamma,\beta)$, ten synthetic target networks were generated, and the above method was applied to each of them to infer a dimensionality $D^*$. Inferred dimensionalities were compared with the original dimension $D$ from which the target synthetic network was generated. Confusion matrices to visualize the performance of the inference method are shown in Fig.~\ref{fig:pipeline}b. Our predictions generate some small confusion with contiguous values of $D$ to the diagonal, especially for low values of $\beta$ and $\gamma$, but this is expected, due to the high heterogeneity of the degree distribution. 

Estimated dimension values for selected real-world target networks using TDA are shown in Table~\ref{tab:summary_dimension}. The resulting values were compared with those obtained by implementing the method described in \cite{almagro2022detecting}, using the configuration space of mean cycle densities $(C_t,C_s,C_p)$,  which have been recalculated using the D-GR procedure.
Inverse values of the congruency index $\rho$ are shown. As additional information, 2D projections of the configuration space ($TP_t$, $TP_s$, $TP_p$) for the selected real-world networks are displayed in Figs.~\textcolor{blue}{S1} and~\textcolor{blue}{S2}. 

\subsection{Dimensionality estimation using neural networks}
\label{subsec:neural_networks}
The method described in the previous section and the approach from~\cite{almagro2022detecting} rely on a large set of surrogate networks. For each new real network, one needs to generate a set of synthetic networks, compute their properties, and use a classifier to detect the dimension. As a consequence, when a new network is considered, the entire pipeline must be repeated. In this section, we propose an alternative based on a neural network that, once trained, can estimate dimensions directly, even for very large networks, generalizing and adapting to new data without retraining.

Neural networks excel for such tasks, since our aim is to approximate an unknown function yielding dimensionality values from a collection of predictors, including mean chordless cycle densities and/or persistence of corresponding filtrations.
For this purpose, we created a database of synthetic complex networks called \textsc{SynNet}, which we use to train a neural network, named \textsc{DimNN}, to estimate the dimensionality of real-world networks. 
The main advantage of this method is that training in our database is performed only once and independently of specific target networks. This avoids the need to generate surrogates for each case under study. 
In total, we produced $792\,000$ synthetic networks generated from the $\mathbb{S}^D$ model (see Section~\ref{sec:synnet} for more details). An 80\%-20\% training-validation split was used.

For each synthetic network in the database, chordless cycle densities were computed as in Section~\ref{subsec:features} and averaged over all edges, as well as total persistence values obtained from the corresponding filtrations, and, additionally, the first moment and the normalized second moment of the degree distribution (i.e., the expected square divided by the square of the expected value). The normalized second moment is sensitive to degree fluctuations and so to very high degree nodes: for highly heterogeneous networks, the ratio becomes large, while for homogeneous networks the ratio is close to~$1$. Hence, the normalized second moment tends to correlate with~$\gamma$. Likewise, the average density of triangles $C_t$ serves as an approximation of the inverse temperature~$\beta$.

Other relevant descriptors of complex networks that we integrated in our feature vectors are minimum and maximum degree, and average neighbor degree, which is denoted by $\langle k_{\rm nn} \rangle$ (see Section~\ref{subsec:features}). This is a measure used to quantify degree-degree correlations in a graph, that is, how the degree of a node relates to the degrees of its neighbors.

\begin{figure*}[ht!]
    \centering
    \includegraphics[width=0.8\textwidth]{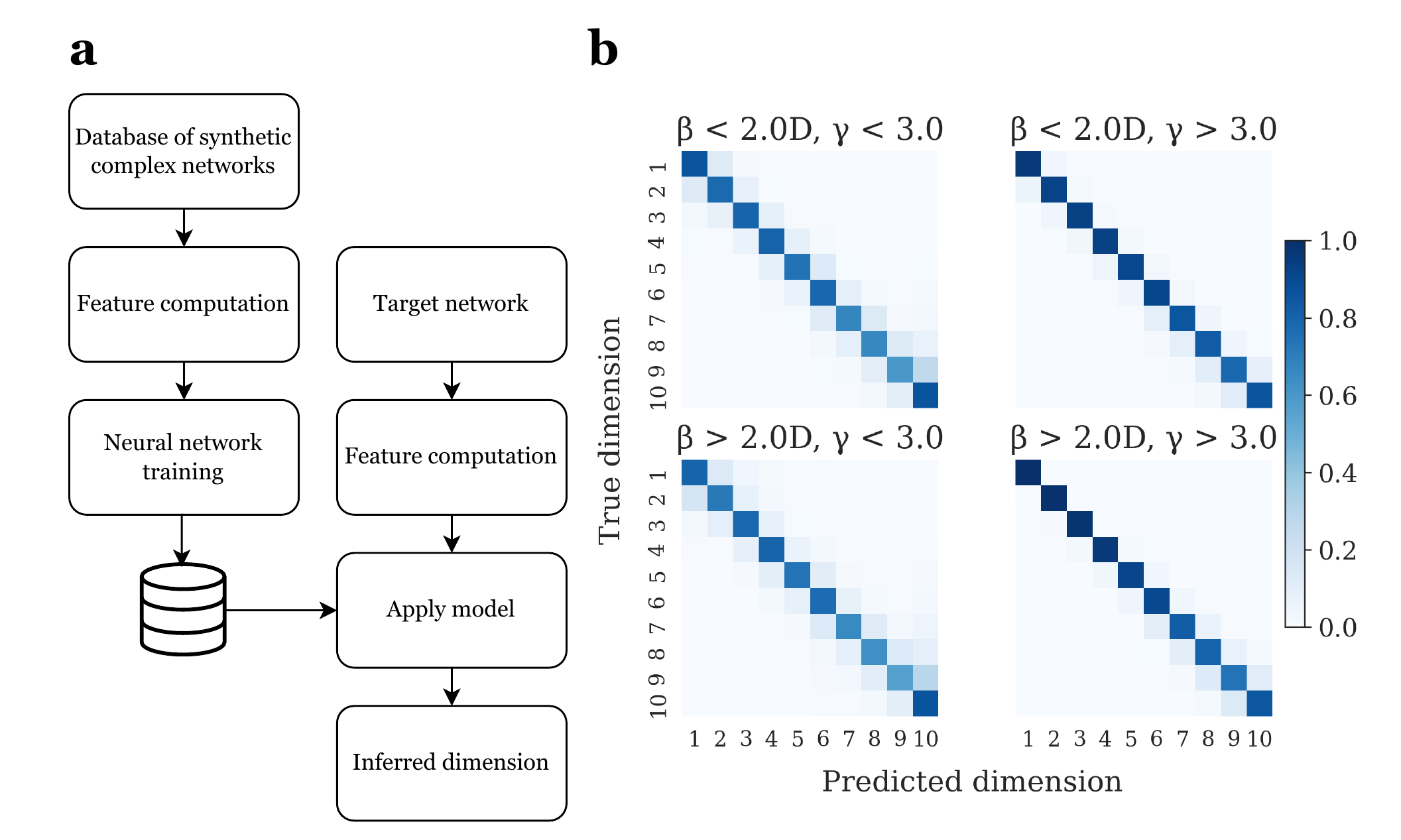}
    \caption{\textbf{Dimensionality estimation using \textsc{DimNN}.} (a) Pipeline of method and (b) confusion matrices. A multilayer perceptron is trained on a database of synthetic complex networks, with features coming from filtrations by densities of chordless cycles concatenated with averages of chordless cycle densities and degree-related graph features. The parameters of the trained model are stored and the model is applied to target networks. Confusion matrices show results for the four combinations of $\beta < 2D$ versus $\beta > 2D$, and $\gamma<3$ versus $\gamma>3$, from a range of values  $1.2\,D$ to $5.0\,D$ for $\beta$ and $2.2$ to $5.0$ for $\gamma$. In each matrix, the rows correspond to the true dimension $D$ and the columns correspond to the inferred dimension~$D^*$.} 
    \label{fig:pipeline2}
\end{figure*}

Feature vectors consisting of number of nodes, average degree, normalized second moment of the degree distribution, minimum and maximum degrees, average neighbor degree, mean chordless cycle densities $C_t$, $C_s$, $C_p$, and total persistences $TP_t$, $TP_s$, $TP_p$ of the corresponding filtrations were fed into a residual multilayer perceptron (see Section~\ref{subsec:database} for details). Our pipeline is shown in Fig.~\ref{fig:pipeline2}a.
The highest classification accuracy obtained with \textsc{DimNN} for estimating the dimensionality in the range $D=1$ to $D=10$ was $83.00\%$ on the validation set with the full feature vector. Confusion matrices are shown in Fig.~\ref{fig:pipeline2}b.

\begin{figure}[ht!]
    \centering
    \includegraphics[width=0.9\linewidth]{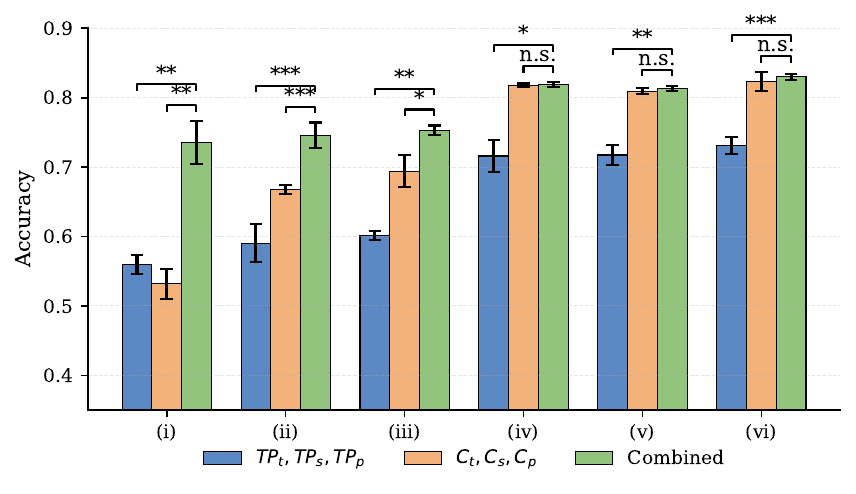}
    \caption{\textbf{Validation accuracies of \textsc{DimNN} using the \textsc{SynNet} database.} 
        Blue color: Accuracies obtained using total persistences $TP_t$, $TP_s$, $TP_p$, plus cumulative features; Orange color: Accuracies using average cycle densities $C_t$, $C_s$, $C_p$, plus cumulative features; Green color: Accuracies obtained combining total persistences and mean cycle densities plus cumulative features.
        Successive columns correspond to incorporating one after the other the following additional features into the model: (i)~no added features; (ii)~number of nodes $N$; (iii)~number of nodes and average degree $\langle k\rangle$; (iv)~number of nodes, average degree, and normalized second moment $\langle k^2\rangle/\langle k\rangle^2$; (v)~number of nodes, average degree, normalized second moment, minimum degree $k_{\rm min}$ and maximum degree $k_{\rm max}$; (vi)~number of nodes, average degree, normalized second moment, minimum degree, maximum degree, and mean average neighbor degree $\langle k_{\rm nn}\rangle$. Results are averaged over five realizations of \textsc{DimNN}, and error bars indicate standard deviation. Horizontal brackets indicate paired comparisons between the combined model and each baseline; significance 
        corresponds to paired $t$-tests with $p<0.05$ (*), $p<0.01$ (**), and $p<0.001$ (***), 
        while ``n.s.'' indicates a non-significant difference.}
    \label{fig:accuracies}
\end{figure}

We also performed an ablation study, selecting subsets of the feature vector for both training and predictions. In Fig.~\ref{fig:accuracies}, we show how the performance changes as a set of features is added at a time. The lowest accuracy is obtained when we use only the vector of mean chordless cycle densities with no other added features. Once degree-related properties are appended to the feature vector ---e.g., minimum, maximum, and average degrees---, the accuracy increases, and incorporating topological information improves the accuracy further. 
In fact, combining total persistence with mean chordless cycle densities maximizes the performance. In Table~\textcolor{blue}{S2}, we show mean validation accuracies as well as the number of epochs and total training times, and a corresponding heatmap is provided in Fig.~\textcolor{blue}{S6}.

Next, we show the results of applying the trained \textsc{DimNN} to detect the dimensionality of real-world complex networks. We analyzed the 10 networks from the previous section and predicted their dimensions. Table~\ref{tab:summary_dimension} compares the inferred dimension across three different methods. One can observe that the predictions do not vary significantly, by only one or two dimensions in most cases, which highlights the robustness of the trained neural network. We incorporated another dataset of real complex networks~\cite{ghasemian2019evaluating}. First, we filtered out bipartite and temporal networks, and restricted to networks with topological properties aligned with the training dataset (Fig.~\textcolor{blue}{S8}). Ranges of \textsc{SynNet} database features are shown in Table~\textcolor{blue}{S1}. In total, we collected 53 new networks, whose properties are summarized in Tables~\textcolor{blue}{S4} and~\textcolor{blue}{S5}.

\begin{figure*}[ht!]
    \centering
    \includegraphics[width=0.9\textwidth]{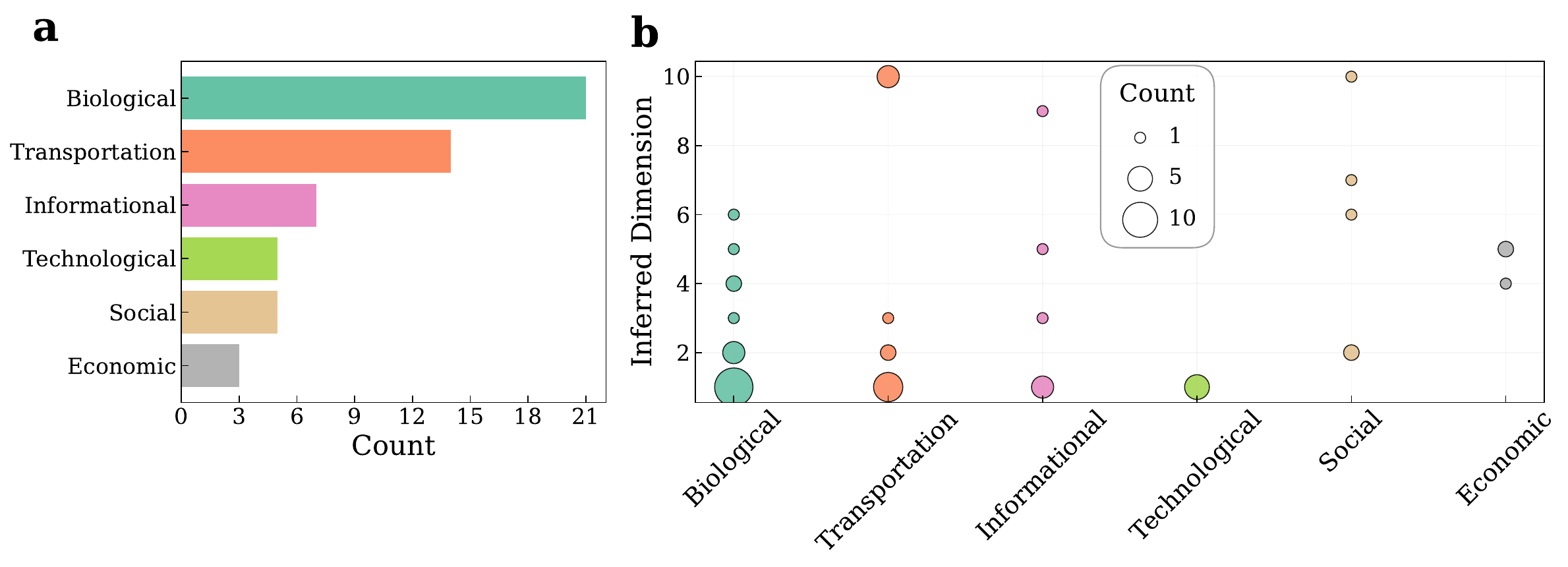}
    \caption{\textbf{Dimensionality estimation of real networks with \textsc{DimNN}.} (a) Number of networks for each network domain. (b) Inferred dimension of real networks using \textsc{DimNN}, grouped by network domain. The marker size indicates the number of networks.} 
    \label{fig:predictions}
\end{figure*}

In Fig.~\ref{fig:predictions}a, we group the dataset by domain, showing that the Biological and Transportation categories together account for more than half of all networks. We then apply our trained neural network to predict the intrinsic dimensions of these real‐world networks (see Fig.~\ref{fig:predictions}b). Biological networks span a wide range of dimensions (from 1 through 6), whereas all networks in the Software domain lie at dimension~1. A~closer look at subdomains (Fig.~\textcolor{blue}{S9}) reveals that web graphs ---an Informational subdomain--- also have dimension~$1$, while language networks (also Informational) appear at dimensions 5 and 9.

Subsequently, to enhance confidence in the predictions made by \textsc{DimNN}, a complementary neural network model was trained. The base architecture was the same as for \textsc{DimNN}, but the loss function was changed to mean squared error (MSE), with the goal of transforming a classification task into a regression task. Thus, contrary to \textsc{DimNN}, the output range of the regressor was not restricted to the interval $[1,10]$.
Since regression yields continuous values, each prediction was rounded to the nearest integer. 
We use the term pseudo-accuracy to denote the accuracy obtained by the regressor after rounding its predictions to integers. Moreover, residual connections were removed.

The regressor model exhibits behavior very consistent with that of \textsc{DimNN} on synthetic networks (Fig.~\textcolor{blue}{S7}). 
The percentage of agreement between the two models on the validation set is 79.33\% if total persistences are combined with mean chordless cycle densities, and reaches 88.86\% when degree-related graph features are added to the training process. 
The discrepancy between the two models is defined as the absolute value of the difference between their predicted dimensions, computed only from those networks in the validation set on which the two models do not agree.
The median discrepancy value was found to be~$1$, with a mean of $1.05$ and a standard deviation of $0.2543$ when a full vector of descriptors was fed into the model. Additional details are given in Table~\textcolor{blue}{S3}.

When applied to the real-world network dataset described in Tables~\textcolor{blue}{S4} and~\textcolor{blue}{S5}, the percentage of agreement between the two models was 64.15\% over 53 networks with full feature vectors in the models, with a median discrepancy of~$1$, a mean of~$2.95$, and a standard deviation of~$3.2032$, indicating the occurrence of a few larger discrepancies.

In addition, we validated the explanatory power of the predicted dimension by generating surrogate networks from hyperbolic embeddings obtained with $D$-Mercator on real networks. Each network was embedded in the dimension predicted by \textsc{DimNN}, and the inferred node positions were used to generate surrogate networks. We then compared multiple topological properties to those of the target networks  
(Figs.~\textcolor{blue}{S10}--\textcolor{blue}{S13}).
Across the tested networks, the surrogates reproduce the degree distribution, clustering spectrum, and $k_{nn}(k)$, and they typically match algebraic connectivity and average shortest-path length, whereas modularity shows weaker agreement, likely reflecting sensitivity to the chosen community detection method. We also compare bond-percolation curves, finding close agreement in the decay of the giant connected component and the inferred percolation threshold 
(Fig.~\textcolor{blue}{S14}).

\section{Discussion}
\label{sec:discussion}
The main contribution of this work is to use densities of cycles (triangles, chordless squares, and chordless pentagons) in a complex network as weights on its edges, and treat each of these three weightings as a filtering function on the underlying graph. Persistence descriptors are computed from the corresponding filtrations and used for estimating the dimensionality of the given network in latent space. 
The expressiveness of edge-based filtrations defined via densities of chordless cycles is further examined in~\cite{vilaminana2026}, where they are compared with other motif-based filtrations in graph isomorphism detection and property prediction tasks. 

Although topological invariants of graphs play a central role in the study of complex networks, methods from topological data analysis have seen limited adoption in this context
\cite{ballester2024on,aktas2019persistence}.  To the best of our knowledge, extended persistent homology of graphs has not previously been applied to dimensionality estimation in complex networks. Ordinary (as opposed to extended) persistent homology is not well-suited for this purpose, since cycles become permanent in graphs. Using Vietoris--Rips simplicial complexes from graph distances is not suitable in our work either, due to the small-world property, and clique complexes of graphs are not optimal either, since triangles become invisible in clique complexes. The extended persistence or lifetime of a cycle in a weighted graph is the difference between the largest weight and the smallest weight of the edges and nodes that form the given cycle. The total lifetime of a set of linearly independent cycles carries relevant information about the geometric structure of the graph.

Other persistence descriptors could be used for analytical purposes. We found persistence of connected components along the filtrations to be less expressive than its cycle-based counterpart.
We also remark that the expressivity of cycle persistence is influenced by certain network parameters, such as inverse temperature ($\beta$). Indeed, in
the point cloud shown in Fig.~\ref{fig:Human2-3D}, low values of $\beta$ tend to separate network dimensions less prominently.

Our results show that total extended persistence computed by means of chordless cycle filtrations yields dimensionality estimates which are comparable with those obtained in previous work \cite{almagro2022detecting}; see Table~\ref{tab:summary_dimension}. Moreover, total persistence improves the accuracy of a neural network classifier if added to feature vectors containing averages of chordless cycle densities and other graph features such as average degree, as shown in Fig.~\ref{fig:accuracies} and in Table~\textcolor{blue}{S2}. Our conclusion is that TDA improves dimensionality estimation using a neural network, even with average cycle densities and degree-related graph descriptors.

In fact, estimating latent network dimensionality without generating surrogates is another major contribution in this article. For this purpose, a universal database of synthetic graphs has been generated from a uniformly distributed range of parameter values. Real-world network dimensionality estimates can then be obtained by running a neural network model trained on our database. This approach considerably reduces the computation time of the estimations and the amount of memory required for this task.
It is far more ambitious than creating an ensemble of surrogates for each given real network under study, as done in the first part of our work.

Fig.~\ref{fig:accuracies} highlights the importance of the normalized second moment $\langle k^2\rangle/\langle k\rangle^2$ in the accuracy of a neural network classifier for latent dimension. This could be due to the fact that the normalized second moment correlates with the exponent~$\gamma$, and knowing $\beta$ and $\gamma$ is crucial for dimensionality estimation. An approximation of $\beta$ is provided by the average density of triangles $C_t$, which is also correlated with the total persistence value~$TP_t$.

The agreement between two independently trained neural network models, each optimized with a different loss function, is a strong indicator of robustness and generalization. When models trained with distinct objective functions (in our case, cross-entropy for classification and mean squared error for regression) converge to similar predictions on previously unseen data, it is unlikely that their performance is due to overfitting specific patterns or uninformative noise in the training set. Rather, such an agreement suggests that both models are approximating the same underlying relationship between the input features (topological and density-based descriptors) and the latent embedding dimension. It is remarkable that a neural network trained on synthetic networks yields results closely aligned with earlier work on the latent dimension of real-world networks.
Moreover, the discrepancies found between the dimensionality predictions of the two models may serve to detect potential out-of-distribution real networks with respect to the D-GR model in the study dataset, which therefore deserve further examination.

Our framework primarily targets geometric networks, i.e., those in the regime $\beta > D$. Recent work, however, suggests that some networks may instead lie in a weakly geometric region~\cite{vanderkolk2022,van2024random}. Extending our approach to this regime and characterizing the interplay between dimensionality and geometricity is a natural direction for future work.

Subsequent research could focus on optimizing the accuracy of a neural network for dimensionality classification of synthetic networks and on improving the performance of latent dimension estimates for real-world networks. 
On the one hand, this could be achieved by enriching the training database with a wider range of complex network shapes. On the other hand, the neural network architecture could be optimized for the intended tasks.
This article only provides a proof of concept for the benefits of a cycle-based filtration in TDA and the feasibility of our deep-learning assisted dimensionality estimation method. While our architecture (Fig.~\textcolor{blue}{S5}) has demonstrated performance consistent with prior work \cite{almagro2022detecting} and has achieved accuracy above $83\%$ on synthetic networks, improving its overall effectiveness remains a challenge.

A further avenue concerns the distribution shift between the synthetic networks used for training and the real networks used for evaluation. In machine-learning terms, this setting can be viewed as a domain adaptation problem, where a predictor is trained on a source domain (synthetic graphs) and deployed on a related but distinct target domain (real graphs)~\cite{farahani2021brief,wang2018deep}. While our current pipeline does not include an explicit adaptation mechanism, a natural extension would be to augment the model with an adversarial domain discriminator that learns to distinguish synthetic from real networks, while the feature extractor is trained to produce domain-invariant representations, potentially improving transfer and robustness under distribution shift~\cite{tzeng2017adversarial}.

\section{Methods}
\label{sec:methods}

\subsection{Multidimensional geometric soft configuration model}
\label{subsec:hyperbolic_models}

\paragraph{The \boldmath$\mathbb{S}^D/\mathbb{H}^{D+1}$ model.}
In the $\mathbb{S}^D$ model~\cite{serrano2008self}, a node $i$ is assigned two hidden variables: a hidden degree $\kappa_i$, quantifying its importance or popularity, and a position $\mathbf{v}_i$ in a $D$-dimensional similarity space, represented as a $D$-sphere. The probability of connection between any pair of nodes $i$ and $j$ follows a gravity law, in which similar nodes are angularly closer and, thus, probably connected~\cite{serrano2021shortest}. Specifically, nodes $i$ and $j$ are connected with probability
\begin{equation} \label{eq:PD}
    p_{ij} = \frac{1}{1+\left( \displaystyle\frac{R \Delta \theta_{ij}}{(\mu \kappa_i \kappa_j)^{1/D}} \right)^\beta}
\end{equation}
where $D$ is the dimension of the model, $\beta$ controls the level of clustering of the network and the coupling of the network with the underlying metric space, $\mu$ controls the average degree, and $\Delta \theta_{ij}$ is the angular distance between nodes $i$ and $j$, which are assigned positions $\mathbf{v}_i$ and $\mathbf{v}_j$ on the $D$-sphere. The radius $R$ of the sphere is set such that the density of $N$ nodes is 1 (without loss of generality). This yields $R = \big[\Gamma \hspace{-0.1cm} \left(\frac{D+1}{2}\right) N/(2 \pi)^{\frac{D+1}{2}} \big]^{1/D}$, where $\Gamma$ is the gamma function~\cite{serrano2021shortest}. For $\beta < D$, networks are unclustered in the infinite-size limit, whereas for $\beta > D$
networks exhibit finite clustering in the thermodynamic limit. Finally, the parameter $\mu$ controls the average degree of the network and is defined as
\begin{align}\label{eq:mu}
    \mu = \frac{\beta\, \Gamma\big(\frac{D}{2}\big) \sin \big(\frac{D\pi}{\beta}\big)}{2 \pi^{1 + \frac{D}{2}} \langle k\rangle}.
\end{align}
The $\mathbb{S}^D$ model is isomorphic to the purely geometric $\mathbb{H}^{D+1}$ model~\cite{budel2024random} in $(D+1)$-hyperbolic space by mapping the hidden degree into the radial coordinates as 
\begin{align}
    {r}_{i}=\hat{R}-\frac{2}{D}\ln \frac{{\kappa }_{i}}{{\kappa }_{0}},\;\;\text{with}\;\;\hat{R}=2\ln \Bigg(\frac{2R}{{(\mu {\kappa }_{0}^{2})}^{1/D}}\Bigg).
\end{align}
A network generation procedure following the $\mathbb{S}^D/\mathbb{H}^{D+1}$ model is described in Algorithm~1 in Supplementary Information, where a cutoff $\kappa_c$ is calculated to prevent excessive fluctuations in the largest expected degrees when $\gamma<3$; see~\cite{boguna2004cut}. 

\paragraph{Microcanonical formulation of $\mathbb{S}^D$ model.}
A microcanonical version of the $\mathbb{S}^1$ model was first proposed in~\cite{starnini2019geometric}, where it was called geometric randomization model (GR). The GR operates on the sequence of observed node degrees and assigns node positions at random in the similarity space. The network is rewired to maximize the likelihood that the new topology is generated by the $\mathbb{S}^1$ model while preserving the observed degrees and, thus, the total number of edges. 

In this work, we extended GR to higher-dimensional similarity spaces, which we call D-GR.  We used D-GR to generate synthetic networks for dimensionality estimation when investigating the relationship between total persistence in homological dimension $1$ computed from three types of chordless cycles filtrations and the inferred dimension (see Section~\ref{subsec:data-driven}). We want to highlight the main difference with respect to the previous approach~\cite{almagro2022detecting}. In~\cite{almagro2022detecting}, the authors infer the set of degrees $\kappa$ from a given network and generate synthetic networks with Eq.~\eqref{eq:PD} given the parameters $\beta$ and~$D$. Although in~\cite{almagro2022detecting} the degree distribution of the generated surrogates is very similar to that of the input network, the D-GR procedure is more constrained and maintains the exact degree values of the input network.

In the D-GR model, we assign to each node $i$ a random position in the $D+1$ dimensional Euclidean space $\mathbf{v}_i \in \mathbb{R}^{D+1}$ with $||\mathbf{v}_i|| = R$. The nodes are uniformly distributed on the $D$-sphere using Marsaglia's algorithm~\cite{marsaglia1972choosing}. 
The rewiring procedure is carried out with the Metropolis--Hastings algorithm, aimed at finding the adjacency matrix that maximizes the likelihood function
\begin{equation}
\label{eq:likelihood}
    \mathcal{L} = \prod_{i<j} p_{ij}^{a_{ij}} [1 - p_{ij}]^{1 - a_{ij}}
\end{equation}
where $p_{ij}$ comes from Eq.~\eqref{eq:PD} and $a_{ij}$ are elements of the adjacency matrix. The method proceeds as described in Algorithm~2 in Supplementary Information.

As shown in \cite{starnini2019geometric}, the probability of swapping links between nodes $i$ and $j$ and between nodes $l$ and $m$ is given by
\begin{equation}\label{eq:lik_rel}
\frac{\mathcal{L}_n}{\mathcal{L}_c} = 
\left( \frac{\Delta \theta_{ij} \, \Delta \theta_{lm}}{\Delta \theta_{il} \, \Delta \theta_{jm}} \right)^{\beta}.
\end{equation}
Notice that Eq.~\eqref{eq:lik_rel} does not depend on the dimension $D$. 

To validate our approach, we generated synthetic networks with known dimension using the $\mathbb{S}^D$ model (see Algorithm~1 in the Supplementary Information), with the following parameters: the number of nodes ($N$) was set to $500$ and the average degree $\langle k \rangle$ to $10$; dimension ($D$) ranging from $1$ to $7$; the inverse temperature parameter ($\beta$) taking values $1.5\,D$ and $2.5\,D$; and the power-law exponent ($\gamma$) taking values $2.7$ and $3.5$. For each set of parameters, we generated 10 network realizations, thus obtaining $280$ synthetic networks. 
For each of these, we produced a set of surrogates using the D-GR method (Algorithm~2 in the Supplementary Information), scanning over different values: dimension ($D$)  ranging from $1$ to $7$ and rescaled inverse temperature $\beta/D$ ranging from $1.2$ to $3.0$ with steps of $0.1$. The geometric randomization was repeated 10 times, yielding $1\,330$ surrogates per synthetic network.
At the end of this procedure, we obtained a total of $372\,400$ networks, that were used in the confusion matrices in Fig.~\ref{fig:pipeline}. 

To infer the dimension of real networks, we proceeded in a similar fashion. We applied the D-GR method with the same set of parameters described above, yielding a total of $1\,330$ surrogates for each real network. Inferred dimensions of real networks are described in Table~\ref{tab:summary_dimension}.

It is worth noting some limitations of this method. The acceptance probability depends on the parameter $\beta$. For very large values of $\beta$, the acceptance ratio becomes binary, i.e., moves that increase the likelihood are almost always accepted, and those that decrease it are almost always rejected, and the likelihood plateau cannot be reached. Thus, the algorithm is restricted to moderate values of $\beta$ and corresponding values of the clustering coefficient. Techniques such as simulated annealing or parallel tempering can restore good mixing in the large $\beta$ regime. In this work, however, we restrict our attention to real networks whose topology is well-captured by a moderate value of $\beta$.

Moreover, even though the $\mathbb{S}^D/\mathbb{H}^{D+1}$ model captures a wide range of topological network properties, some real networks may lie outside the range of values and are located further away from the surrogate networks in the persistence ($TP_t, TP_s, TP_p$) and mean cycle density ($C_t, C_s, C_p$) configuration spaces.

\subsection{Graph features}
\label{subsec:features}
\paragraph{Densities of chordless cycles.}
\label{par:chordless_cycle_filtrations}
Let $G=(V,E)$ be a graph and $e_{ij}=\{v_i,v_j\}$ be an edge between nodes $v_i$ and $v_j$ in~$E$, with degrees $k_i>1$ and~$k_j>1$, respectively. The density of triangles corresponding to the edge~$e_{ij}$, also called edge clustering coefficient \cite{serrano2006clustering}, is the number $\#\triangle_{ij}$ of edge triangles in $G$ containing $e_{ij}$ divided by the maximum possible number of triangles in $G$ containing~$e_{ij}$ given the degrees $k_i$ and~$k_j$, that~is,
\begin{equation}\label{eq: density_triangles}
    C_t(e_{ij}) = \frac{\# \triangle_{ij}}{\min (k_i, k_j) - 1}.
\end{equation}
An edge cycle is chordless if there is no edge between its nodes except those that form the cycle.
Density of squares, denoted $C_s(e_{ij})$, is defined by dividing the number of chordless edge squares in $G$ containing $e_{ij}$ by the maximum possible number of such squares given the degrees $k_i$ and $k_j$ and the existing triangles through~$e_{ij}$ in~$G$.
To define the density of pentagons $C_p(e_{ij})$ per edge $e_{ij}$, we count the number of chordless pentagons containing $e_{ij}$  and normalize it by the maximum possible number of such pentagons, assuming known the degrees of $v_i$ and $v_j$ and the degrees of their respective neighbors.
One could also define similar densities per node. However, in~\cite{serrano2006clustering}, the authors found the definitions relative to edges to be more stable with respect to degree heterogeneity than those relative to nodes.

Regarding the computational cost of computing chordless cycles, the time complexities for computing densities of triangles, squares, and pentagons are $O(N \langle k \rangle^2)$, $O(N\langle k \rangle^3)$, and $O(N\langle k \rangle^4)$, respectively, where $\langle k\rangle$ is the average degree and $N=\#E$ is the number of edges. In sparse graphs, $\langle k \rangle \ll N$. Although the computational cost for large graphs may be high, the computations can be easily parallelized or implemented on a GPU using CUDA~\cite{cuda}.

\paragraph{Extended persistent homology.}
\label{par:topological_descriptors}
In this work, we use persistent homology to extract information from a graph $G$ equipped with a filtration $\{G_t\}$, where $t$ is a real-valued parameter. Persistence refers to the evolution of cycles along the values of the given filtration.

Homology of graphs is a special case of simplicial homology of simplicial complexes~\cite{hatcher}. In the case of graphs, homology is determined by two numbers, called Betti numbers, namely the number $\beta_0$ of connected components and the cardinality $\beta_1$ of a maximal set of linearly independent cycles. Cycles are finite formal sums $z=\sum \lambda_ie_i$ of oriented edges with coefficients $\lambda_i=\pm 1$ such that $\sum \lambda_i \partial e_i=0$, where, for an edge $e\in E$ from $v_0$ to $v_1$, we denote $\partial e=v_1-v_0$. Thus, cycles are algebraic representations of closed edge paths in the given graph.

In a filtered simplicial complex $\{K_t\}$, the birth of a cycle $z$ of any dimension is the $t$-value at which $z$ appears in $K_t$, and the death of $z$ is the $t$-value at which $z$ becomes the boundary of a higher chain. Hence, in the case of a filtered graph, the birth of a cycle $z$ is the $t$\nobreakdash-value at which $z$ is formed, and the death value is infinite, since a graph does not contain higher simplices. 

To avoid infinite persistence values, we use extended persistence, as in~\cite{CS-E-H2009,carriere2019perslay}. For a graph $G=(V,E)$ equipped with a node weighting $w_V\colon V\to\mathbb{R}$, the sublevel filtration $\{G_t\}$ is defined as $G_t=(V_t,E_t)$, where $V_t=\{v\in V\mid w_V(v)\le t\}$ and $E_t$ is the subset of $E$ spanned by~$V_t$, and the superlevel filtration $\{G^t\}$ is defined as $G^t=(V^t,E^t)$, where $V^t=\{v\in V\mid w_V(v)\ge t\}$ and $E^t$ is the subset of $E$ spanned by~$V^t$. Then the extended persistence of a cycle $z$ in $G$ is defined as $|d-b|$, where $b$ is the birth value of $z$ in the sublevel filtration $\{G_t\}$ and $d$ is the birth value of $z$ in the superlevel filtration $\{G^t\}$. Note, however, that no edge weighting can be defined on $G$ compatibly with $w_V$ which is consistent with both $\{G_t\}$ and $\{G^t\}$, in general.

In our work, chordless cycle filtrations are defined by means of edge weightings, not node weightings. 
Although it is perfectly possible to exchange the roles of nodes and edges in the definitions of sublevel and superlevel filtrations, extended persistence calculations have been carried out using the Python library Gudhi~\cite{gudhi}, which only provides software for computing extended persistence of graphs with a weighting on their nodes (see Section~2.1 of \cite{carriere2019perslay}).
Therefore, we introduce a method to replace a graph $G=(V,E)$ equipped with an edge weighting $w_E\colon E\to\mathbb{R}$ by a topologically equivalent graph $G'=(V',E')$ in which $E\subseteq E'$ and endowed with a node weighting $w_{V'}\colon V'\to\mathbb{R}$ yielding the same extended persistence diagram as~$G$. 

To achieve this, we split each node $v_i\in V$ of degree $k_i$ into $k_i$ distinct nodes $v_{i,j}\in V'$, where $j$ ranges over the subindices of the neighbors of~$v_i$. Each node $v_{i,j}$ is assigned weight $w_{V'}(v_{i,j})=w_E(e_{ij})$ where $e_{ij}=\{v_i,v_j\}$.
The set $E'$ contains horizontal edges $\{v_{i,j},v_{j,i}\}$ for all $i,j$, joining nodes with the same weight, and vertical edges, connecting each string of nodes $v_{i,j}$ with a fixed $i$ value, sequentially from smallest weight to largest, as in Fig.~\ref{fig:blow-up}. Hence, the set of horizontal edges in $E'$ is in bijective correspondence with~$E$, and the weighting $w_{V'}$ is consistent with $w_E$ on all edges of~$E$, in both the sublevel and superlevel filtrations of~$G'$.

We call $G'$ a degree-splitting subdivision of $G$.
An example is shown in Fig.~\ref{fig:blow-up}, and a corresponding sequence of sublevel graphs and superlevel graphs explaining the choice of generating cycles for extended persistence is depicted in Fig.~\textcolor{blue}{S4}.

The graph $G'$ is then fed into Gudhi \cite{gudhi} as a node-weighted graph, and total persistence $\sum|d_i-b_i|$ of cycles in the corresponding extended persistence diagram is recorded (relative cycles provided by the Gudhi software, if any, are discarded). The graph $G'$ has the same Betti numbers as $G$ at the same thresholds, since $G$ is obtained from $G'$ by contracting vertical edges, which does not change homology.

\paragraph{Degree-related graph features.}
\label{par:other_features}

The following descriptors from complex network theory are calculated to train the models used in Section~\ref{subsec:neural_networks}.
For each graph $G=(V,E)$, we consider the average degree of its nodes, $\langle k\rangle=(1/\#V)\sum_{v\in V}\deg(v)$. The minimal degree $k_{\rm min}$ and maximal degree $k_{\rm max}$ are also considered. 
The average neighbor degree of $G$ is the average of the mean neighbor degree of its nodes:
\[
\langle k_{\rm nn} \rangle=\frac{1}{\#V}\sum_{v\in V}\Bigg(\frac{1}{\deg(v)}\sum_{u\sim v}\deg(u)\Bigg).
\]

The normalized second moment of $G$ is defined as the quotient $\langle k^2\rangle/\langle k\rangle^2$, where we denote $\langle k^2\rangle=(1/\#V)\sum_{v\in V}\deg(v)^2$. Its significance is due to the fact that it correlates inversely with the power-law exponent $\gamma$ in scale-free networks~\cite{Albert2002}.

\subsection{Database and neural network architecture}
\label{subsec:database}

For better understanding and comparison of the results, we used the same real-world networks as in~\cite{almagro2022detecting}. These are undirected networks with fewer than 100 000 nodes from very different domains. Tables~\textcolor{blue}{S4} and \textcolor{blue}{S5} include a detailed description of the data as reported in~\cite{almagro2022detecting}.

\paragraph{\textsc{SynNet} dataset for inferring dimension using neural networks.}\label{sec:synnet}
Deep networks require large amounts of data to avoid overfitting and to learn robust features, due to their large number of parameters. In fact, neural networks generally require large datasets to achieve high performance~\cite{krizhevsky2012imagenet,lecun2015deep}. However, with the $\mathbb{S}^D$ model, we tackle this challenge by generating plenty of synthetic graphs with known dimensions to train neural networks.

We prepared a dataset of $792\,000$ synthetic networks generated using the $\mathbb{S}^D$ model by means of the method described in Algorithm~1 in the Supplementary Information with the following input parameters: dimension ($D$) from $1$ to $10$; number of nodes ($N$) with values $200$, $400$, $750$, $1000$, $2500$; power-law exponent ($\gamma$) with values $2.2$, $2.4$, $3.0$, $4.0$, $5.0$; average degree ($\langle k\rangle$) with values 4, 8, 12, 25; and rescaled inverse temperature ($\beta/D$) with values $1.2$, $1.4$, $1.6$, $1.8$, $2.0$, $2.2$, $2.5$, $2.8$, $3.0$, $3.5$, $4.0$, $5.0$. For each set of parameters, $66$ network realizations were made. The dataset covers a wide range of network properties ---in particular, $\langle k_{nn} \rangle \in [4.4, 300.8]$, $C_t \in [0.073, 0.891]$, $C_s \in [0.00505, 0.24193]$, $C_p = [0.00010, 0.02165]$.

The database contains a set of feature vectors for each network, aiming for its use in a multilayer perceptron (MLP), which is described in Section~\ref{par:neural_networks}.
The following descriptors were computed, resulting in a $12$-dimensional feature vector:
$N$, $\langle k \rangle$, $\langle k^2\rangle/\langle k\rangle^2$, $k_{\rm min}$, $k_{\rm max}$, $\langle k_{\rm nn} \rangle$, $C_t$, $C_s$, $C_p$, $TP_t$, $TP_s$, $TP_p$, as described in Section~\ref{subsec:features}.
Once a MLP is trained, the same features are computed for real networks in order to predict their dimensions. 

It is worth mentioning that some real networks could have very distinct topological properties that are not included in our dataset. Hence, neural networks might be prone to misclassifying them. We can overcome this issue by extending the range or parameters used to generate the dataset. In the released code, we provide a check for the out-of-distribution parameters.

\paragraph{Neural network \textsc{DimNN} model.}
\label{par:neural_networks}
For the classification task, a deep multilayer perceptron (MLP) architecture combined with the techniques of residual networks (ResNets) was used, as in \cite{Touvron2021}. Deeper neural networks are capable of capturing highly non-linear relationships in data, but training very deep architectures often leads to optimization difficulties such as vanishing gradients. ResNets address these issues through the introduction of skip (residual) connections, which add the input of a layer to its output. Such mappings facilitate gradient flow and stabilize training, enabling the construction of deeper and more expressive models \cite{resnet}.

The proposed architecture integrates residual connections into a fully connected MLP, allowing the model to benefit from the depth of representation while maintaining stable training dynamics. The network comprises 21 hidden layers with ReLu activations, and the configuration depicted in Fig.~\textcolor{blue}{S5}. The output layer uses a softmax activation. A dropout regularization of $50\%$ is applied throughout the network to mitigate overfitting and enhance generalization. The training process was carried out using AdamW as optimizer, with a learning rate of $0.0005$ and incorporating early stopping to prevent overfitting.
This architecture benefits from the expressive power of deep learning while incorporating the robustness of residual learning, making it convenient for the proposed classification task.

\section*{Data Availability}
The real network datasets used in this study are available from the sources referenced in the manuscript and the Supplementary Information. 
The \textsc{SynNet} dataset of $792\,000$ synthetic networks generated with the $\mathbb{S}^D$ model used for neural network training, along with the neural network checkpoints, is available on Zenodo~\cite{zenodo}.

\section*{Code Availability}
The open-source code generated in this work, along with the code to reproduce the figures, is available on GitHub at 
\url{https://github.com/networkgeometry/detecting-dimensionality-TDA-DimNN}.

\section*{Acknowledgments}
A.F.M.\ was supported by MCIN/AEI under grant PRE2020-094372; A.F.M.\ and C.C.\ were partially funded through MCIN/AEI
grants PID2020-117971GB-22 and PID2022-136436NB-I00, as well as AGAUR grant 2021 SGR 00697; R.J.\ acknowledges support from the fellowship FI-SDUR funded by Generalitat de Catalunya; M.A.S. acknowledges support from grant no.\ TED2021-129791B-I00 funded by MCIN/AEI/10.13039/501100011033 and by \emph{European Union NextGenerationEU/PRTR}, and grant no.\ PID2022-137505NB-C22 funded by MCIN/AEI/10.13039/501100011033 and by ERDF's \emph{A way of making Europe}.

\section*{Author Contributions Statement}
M.A.S., M.V.M.\ and A.F.M.\ designed research; M.A.S.\ and R.J.\ contributed network science methods; C.C., A.F.M.\ and M.V.M.\ contributed persistent homology methods; A.F.M.\ and R.J.\ performed the computations and numerical simulations; A.F.M.\ generated the \textsc{DimNN} code; R.J.\ generated the \textsc{SynNet} dataset. All authors conducted research, analyzed the results, discussed implications, and contributed to the writing of the manuscript.

\section*{Competing Interests Statement}
The authors declare no competing interests.

\bibliography{references}

@article{tillquist2023getting,
  title={Getting the lay of the land in discrete space: A survey of metric dimension and its applications},
  author={Tillquist, Richard C. and Frongillo, Rafael M. and Lladser, Manuel E.},
  journal={SIAM Review},
  volume={65},
  number={4},
  pages={919--962},
  year={2023},
  publisher={SIAM}
}

@article{lacasa2013correlation,
  title={Correlation dimension of complex networks},
  author={Lacasa, Lucas and G{\'o}mez-Gardenes, Jes{\'u}s},
  journal={Physical Review Letters},
  volume={110},
  number={16},
  pages={168703},
  year={2013},
  publisher={APS}
}

@article{daqing2011dimension,
  title={Dimension of spatially embedded networks},
  author={Daqing, Li and Kosmidis, Kosmas and Bunde, Armin and Havlin, Shlomo},
  journal={Nature Physics},
  volume={7},
  number={6},
  pages={481--484},
  year={2011},
  publisher={Nature Publishing Group UK London}
}

@article{kim2007box,
  title={A box-covering algorithm for fractal scaling in scale-free networks},
  author={Kim, J.~S. and Goh, K.-I. and Kahng, B. and Kim, D.},
  journal={Chaos: An Interdisciplinary Journal of Nonlinear Science},
  volume={17},
  number={2},
  year={2007},
  publisher={AIP Publishing}
}

@article{wei2014new,
  title={A new information dimension of complex networks},
  author={Wei, Daijun and Wei, Bo and Hu, Yong and Zhang, Haixin and Deng, Yong},
  journal={Physics Letters A},
  volume={378},
  number={16-17},
  pages={1091--1094},
  year={2014},
  publisher={Elsevier}
}

@article{shanker2007defining,
  title={Defining dimension of a complex network},
  author={Shanker, O.},
  journal={Modern Physics Letters B},
  volume={21},
  number={06},
  pages={321--326},
  year={2007},
  publisher={World Scientific}
}

@article{song2007calculate,
  title={How to calculate the fractal dimension of a complex network: the box covering algorithm},
  author={Song, Chaoming and Gallos, Lazaros K. and Havlin, Shlomo and Makse, Hern{\'a}n A.},
  journal={Journal of Statistical Mechanics: Theory and Experiment},
  volume={2007},
  number={03},
  pages={P03006--P03006},
  year={2007}
}

@article{eguiluz2003effective,
  title={Effective dimensions and percolation in hierarchically structured scale-free networks},
  author={Egu{\'\i}luz, V{\'\i}ctor M and Hern{\'a}ndez-Garc{\'\i}a, Emilio and Piro, Oreste and Klemm, Konstantin},
  journal={Physical Review E},
  volume={68},
  number={5},
  pages={055102},
  year={2003},
  publisher={APS}
}

@inproceedings{avrachenkov2019eigenvalues,
  title={Eigenvalues and spectral dimension of random geometric graphs in thermodynamic regime},
  author={Avrachenkov, Konstantin and Cottatellucci, Laura and Hamidouche, Mounia},
  booktitle={International Conference on Complex Networks and Their Applications},
  pages={965--975},
  year={2019},
  organization={Springer}
}

@article{PhysRevLett.76.1091,
  title = {Universal Properties of Spectral Dimension},
  author = {Burioni, Raffaella and Cassi, Davide},
  journal = {Physical Review Letters},
  volume = {76},
  issue = {7},
  pages = {1091--1093},
  numpages = {0},
  year = {1996},
  month = {Feb},
  publisher = {American Physical Society},
  doi = {10.1103/PhysRevLett.76.1091},
  url = {https://link.aps.org/doi/10.1103/PhysRevLett.76.1091}
}

@article{jankowski2023d,
  title={The {$D$-Mercator} method for the multidimensional hyperbolic embedding of real networks},
  author={Jankowski, Robert and Allard, Antoine and Bogu{\~n}{\'a}, Mari{\'a}n and Serrano, M. {\'A}ngeles},
  journal={Nature Communications},
  volume={14},
  number={1},
  pages={7585},
  year={2023},
  publisher={Nature Publishing Group UK London}
}

@article{almagro2022detecting,
    title={Detecting the ultra low dimensionality of real networks},
    author={Almagro, P. and Bogu{\~n}{\'a}, M. and Serrano, M. \'A.},
    journal={Nature Communications},
    volume={13},
    pages={6096},
    year={2022},
    month={Oct},
    publisher={Nature Publishing Group}
}

@article{aktas2019persistence,
    title={Persistence homology of networks: Methods and applications},
    author={Aktas, M. E. and Akbas, E. and El Fatmaoui, A.},
    journal={Applied Network Science},
    volume={4},
    number={61},
    year={2019},
    publisher={Springer}
}

@article{hernandez2020simplicial,
    title={Simplicial degree in complex networks. {A}pplications of topological data analysis to network science},
    author={Hern{\'a}ndez Serrano, Daniel and Hern{\'a}ndez Serrano, Juan and S{\'a}nchez G{\'o}mez, Dar{\'i}o},
    journal={Chaos, Solitons and Fractals},
    volume={137},
    year={2020},
    month={Aug},
    pages={109839},
    publisher={Elsevier}
}

@inproceedings{carriere2019perslay,
    title={{PersLay: A neural network layer for persistence diagrams and new graph topological signatures}},
    author={Carri{\`e}re, M. and Chazal, F. and Ike, Y. and Lacombe, T. and Royer, M. and Umeda, Y.},
    booktitle={Proceedings of the 23rd International Conference on Artificial
    Intelligence and Statistics (AISTATS)},
    series={PMLR},
    volume={108},
    year={2020}
}

@article{krioukov2010hyperbolic,
    title={Hyperbolic geometry of complex networks},
    author={Krioukov, D. and Papadopoulos, F. and Kitsak, M. and Vahdat, A. and Bogu{\~n}{\'a}, M.},
    journal={Physical Review E -- Statistical, Nonlinear, and Soft Matter Physics},
    volume={82},
    number={9},
    year={2010},
    month={Sep},
    publisher={American Physical Society}
}

@book{serrano2021shortest,
    title={The Shortest Path to Network Geometry},
    author={Serrano, M. \'A. and Bogu{\~n}{\'a}, M.},
    publisher={Cambridge University Press},
    address={Cambridge, UK},
    year={2021}
}

@article{budel2024random,
  title={Random hyperbolic graphs in $d+1$ dimensions},
  author={Budel, Gabriel and Kitsak, Maksim and Aldecoa, Rodrigo and Zuev, Konstantin and Krioukov, Dmitri},
  journal={Physical Review E},
  volume={109},
  number={5},
  pages={054131},
  year={2024},
  publisher={APS}
}

@article{starnini2019geometric,
  title={Geometric randomization of real networks with prescribed degree sequence},
  author={Starnini, Michele and Ortiz, Elisenda and Serrano, M. {\'A}ngeles},
  journal={New Journal of Physics},
  volume={21},
  number={5},
  pages={053039},
  year={2019},
  publisher={IOP Publishing}
}

@article{marsaglia1972choosing,
  title={Choosing a point from the surface of a sphere},
  author={Marsaglia, George},
  journal={The Annals of Mathematical Statistics},
  volume={43},
  number={2},
  pages={645--646},
  year={1972},
  publisher={Institute of Mathematical Statistics}
}

@article{wasserman2018topological,
  title={Topological data analysis},
  author={Wasserman, Larry},
  journal={Annual Review of Statistics and its Application},
  volume={5},
  pages={501--532},
  year={2018},
  publisher={Annual Reviews}
}

@book{carlsson2021topological,
  title={Topological data analysis with applications},
  author={Carlsson, Gunnar and Vejdemo-Johansson, Mikael},
  year={2021},
  publisher={Cambridge University Press}
}

@article{forman2003bochner,
  title={Bochner's method for cell complexes and combinatorial {R}icci curvature},
  author={Robin Forman},
  journal={Discrete \& Computational Geometry},
  volume={29},
  pages={323--374},
  year={2003},
  publisher={Springer}
}

@article{freeman1977set,
  title={A set of measures of centrality based on betweenness},
  author={Freeman, Linton C.},
  journal={Sociometry},
  volume={40},
  number={10},
  pages={35--41},
  year={1977},
}

@incollection{bobrowski2022random,
  title={Random simplicial complexes: {M}odels and phenomena},
  author={Bobrowski, Omer and Krioukov, Dmitri},
  booktitle={Higher-Order Systems},
  pages={59--96},
  year={2022},
  publisher={Springer}
}

@article{salnikov2018simplicial,
  title={Simplicial complexes and complex systems},
  author={Salnikov, Vsevolod and Cassese, Daniele and Lambiotte, Renaud},
  journal={European Journal of Physics},
  volume={40},
  number={1},
  pages={014001},
  year={2018},
  publisher={IOP Publishing}
}

@book{bianconi2021higher,
  title={{Higher-Order Networks}},
  author={Bianconi, Ginestra},
  year={2021},
  publisher={Cambridge University Press}
}

@article{torres2020simplicial,
  title={Simplicial complexes: {H}igher-order spectral dimension and dynamics},
  author={Torres, Joaqu{\'\i}n J. and Bianconi, Ginestra},
  journal={Journal of Physics: Complexity},
  volume={1},
  number={1},
  pages={015002},
  year={2020},
  publisher={IOP Publishing}
}

@article{sizemore2019importance,
  title={The importance of the whole: {T}opological data analysis for the network neuroscientist},
  author={Sizemore, Ann E. and Phillips-Cremins, Jennifer E. and Ghrist, Robert and Bassett, Danielle S.},
  journal={Network Neuroscience},
  volume={3},
  number={3},
  pages={656--673},
  year={2019}
}

@article{taylor2015topological,
  title={Topological data analysis of contagion maps for examining spreading processes on networks},
  author={Taylor, Dane and others},
  journal={Nature Communications},
  volume={6},
  number={1},
  pages={7723},
  year={2015},
  publisher={Nature Publishing Group UK London}
}

@article{boguna2021network,
  title={Network geometry},
  author={Bogu{\~n}{\'a}, Mari{\'a}n and Bonamassa, Ivan and De Domenico, Manlio and Havlin, Shlomo and Krioukov, Dmitri and Serrano, M. {\'A}ngeles},
  journal={Nature Reviews Physics},
  volume={3},
  number={2},
  pages={114--135},
  year={2021},
  publisher={Nature Publishing Group UK London}
}

@article{serrano2008self,
  title={Self-similarity of complex networks and hidden metric spaces},
  author={Serrano, M. {\'A}ngeles and Krioukov, Dmitri and Bogu{\~n}{\'a}, Mari{\'a}n},
  journal={Physical Review Letters},
  volume={100},
  number={7},
  pages={078701},
  year={2008},
  publisher={APS}
}

@inproceedings{krizhevsky2012imagenet,
  title     = {ImageNet Classification with Deep Convolutional Neural Networks},
  author    = {Krizhevsky, Alex and Sutskever, Ilya and Hinton, Geoffrey E.},
  booktitle = {Advances in Neural Information Processing Systems},
  pages     = {1097--1105},
  year      = {2012}
}

@article{lecun2015deep,
  title     = {Deep Learning},
  author    = {LeCun, Yann and Bengio, Yoshua and Hinton, Geoffrey},
  journal   = {Nature},
  volume    = {521},
  number    = {7553},
  pages     = {436--444},
  year      = {2015},
  publisher = {Nature Publishing Group}
}

@article{serrano2006clustering,
  title={Clustering in complex networks. {I.} {G}eneral formalism},
  author={Serrano, M. {\'A}ngeles and Bogu{\~n}{\'a}, Mari{\'a}n},
  journal={Physical Review E -- Statistical, Nonlinear, and Soft Matter Physics},
  volume={74},
  number={5},
  pages={056114},
  year={2006},
  publisher={APS}
}

@article{boguna2004cut,
  title={Cut-offs and finite size effects in scale-free networks},
  author={Bogu{\~n}{\'a}, Mari{\'a}n and Pastor-Satorras, Romualdo and Vespignani, Alessandro},
  journal={The European Physical Journal B},
  volume={38},
  pages={205--209},
  year={2004},
  publisher={Springer}
}

@inproceedings{resnet,
  author={He, Kaiming and Zhang, Xiangyu and Ren, Shaoqing and Sun, Jian},
  booktitle={2016 IEEE Conference on Computer Vision and Pattern Recognition (CVPR)}, 
  title={Deep Residual Learning for Image Recognition}, 
  year={2016},
  volume={},
  number={},
  pages={770-778},
  doi={10.1109/CVPR.2016.90}
}

@book{gudhi
, title        = "{GUDHI} User and Reference Manual"
, author      = "{The GUDHI Project}"
, publisher     = "{GUDHI Editorial Board}"
, year         = 2015
, url =    "http://gudhi.gforge.inria.fr/doc/latest/"
}

@inproceedings{
ballester2024on,
title={On the Expressivity of Persistent Homology in Graph Learning},
author={Rub{\'e}n Ballester and Bastian Rieck},
booktitle={The Third Learning on Graphs Conference},
year={2024},
url={https://openreview.net/forum?id=phs5A7bUPA}
}

@inproceedings{cuda,
author = {Nickolls, John and Buck, Ian and Garland, Michael and Skadron, Kevin},
title = {Scalable parallel programming with {CUDA}},
year = {2008},
isbn = {9781450378451},
publisher = {Association for Computing Machinery},
address = {New York, NY, USA},
url = {https://doi.org/10.1145/1401132.1401152},
doi = {10.1145/1401132.1401152},
booktitle = {ACM SIGGRAPH 2008 Classes},
articleno = {16},
numpages = {14},
location = {Los Angeles, California},
series = {SIGGRAPH '08}
}

@book{Hatcher,
title        = {Algebraic Topology},
author      = {Hatcher, Allen},
publisher     = {Cambridge University Press},
year         = {2002}
}

@article{Horak2009,
title     = {Persistent homology of complex networks},
  author    = {Danijela Horak and Slobodan Maleti\'c and Milan Rajkovi\'c},
  journal   = { J. Stat. Mech.},
  volume    = {2009},
  number    = {},
  pages     = {P03034},
  year      = {2009},
  publisher = {}
}

@article{Myers2019,
title     = {Persistent homology of complex networks for dynamic state detection},
  author    = {Audun Myers and Elizabeth Munch and Firas A. Khasawneh},
  journal   = {Physical Review E},
  volume    = {100},
  number    = {022314},
  pages     = {},
  year      = {2019},
  publisher = {}
}

@article{Touvron2021,
  author={Touvron, Hugo and others},
  journal={IEEE Transactions on Pattern Analysis and Machine Intelligence}, 
  title={{ResMLP}: Feedforward networks for image classification with data-efficient training}, 
  year={2023},
  volume={45},
  number={4},
  pages={5314-5321},
  doi={10.1109/TPAMI.2022.3206148}}

@article{giusti2015,
title     = {Clique topology reveals intrinsic geometric structure in neural correlations},
  author    = {Chad Giusti and Eva Pastalkova and Carina Curto and Vladimir Itsko},
  journal   = {PNAS},
  volume    = {112},
  number    = {44},
  pages     = {13455--13460},
  year      = {2015},
  publisher = {}
}

@article{das2023,
title     = {Topological data analysis of human brain networks through order statistics},
  author    = {Soumya Das and D. Vijay Anand and Moo K. Chung},
  journal   = {PLoS One},
  volume    = {18},
  number    = {3},
  pages     = {e0276419},
  year      = {2023},
  publisher = {}
}

@article{guerra2021,
title     = {Homological scaffold via minimal homology bases},
  author    = {Marco Guerra and Alessandro De Gregorio and Ulderico Fugacci and Giovanni Petri and Francesco Vaccarino},
  journal   = {Scientific Reports},
  volume    = {11},
  number    = {5355},
  pages     = {},
  year      = {2021},
  publisher = {}
}

@article{petri2014,
title     = {Homological scaffolds of brain functional networks},
  author    = {G. Petri and P. Expert and F. Turkheimer and R. Carhart-Harris and D. Nutt and P. J. Hellyer and F. Vaccarino},
  journal   = {Journal of the Royal Society Interface},
  volume    = {11},
  number    = {101},
  pages     = {},
  year      = {2014},
  publisher = {}
}

@article{jhun2022,
title     = {Topological analysis of the latent geometry of a complex network},
  author    = {Bukyoung Jhun},
  journal   = {Chaos},
  volume    = {32},
  number    = {013116},
  pages     = {},
  year      = {2022},
  publisher = {}
}

@article{kannan2019,
title     = {Persistent homology of unweighted complex networks via discrete {M}orse theory},
  author    = {Harish Kannan and Emil Saucan and Indrava Roy and  Areejit Samal},
  journal   = {Scientific Reports},
  volume    = {9},
  number    = {13817},
  pages     = {},
  year      = {2019},
  publisher = {}
}

@article{Albert2002,
title     = {Statistical mechanics of complex networks},
  author    = {R\'eka Albert and Albert-L\'aszl\'o Barab\'asi},
  journal   = {Reviews of Modern Physics},
  volume    = {74},
  number    = {47},
  pages     = {},
  year      = {2002},
  publisher = {}
}

@article{ghasemian2019evaluating,
  title={Evaluating overfit and underfit in models of network community structure},
  author={Ghasemian, Amir and Hosseinmardi, Homa and Clauset, Aaron},
  journal={IEEE Transactions on Knowledge and Data Engineering},
  volume={32},
  number={9},
  pages={1722--1735},
  year={2019},
  publisher={IEEE}
}

@book{edelsbrunner2022,
  title={Computational Topology: An Introduction},
  author={Edelsbrunner, Herbert and Harer, John L.},
  year={2022},
  publisher={American Mathematical Soceity}
}

@article{CS-E-H2009,
  title={Extending persistence Using {P}oincar\'e and {L}efschetz duality},
  author={Cohen-Steiner, David and Edelsbrunner, Herbert and Harer, John},
  journal={Foundations of Computational Mathematics},
  volume={9},
  number={1},
  pages={79--103},
  year={2009}
}

@article{ising1925,
  title        = {Beitrag zur {T}heorie des {F}erromagnetismus},
  author       = {Ising, Ernst},
  journal      = {Zeitschrift f{\"u}r Physik},
  volume       = {31},
  pages        = {253--258},
  year         = {1925},
  doi          = {10.1007/BF02980577}
}

@article{onsager1944,
  title={Crystal statistics. {I}. {A} two-dimensional model with an order-disorder transition},
  author={Onsager, Lars},
  journal={Physical Review},
  volume={65},
  number={3-4},
  pages={117},
  year={1944},
  publisher={APS}
}

@book{redner2001,
  title        = {A Guide to First-Passage Processes},
  author       = {Redner, Sidney},
  publisher    = {Cambridge University Press},
  year         = {2001},
  isbn         = {9780521652483}
}

@inproceedings{dvoretzky1951,
  title={Some problems on random walk in space},
  author={Dvoretzky, Aryeh and Erd\H{o}s, Paul},
  booktitle={Proc. Second Berkeley Symp. Math. Statist. Probab},
  pages={353--367},
  year={1951}
}

@article{condamin2007,
  title        = {First-passage times in complex scale-invariant media},
  author       = {Condamin, St{\'e}phane and B{\'e}nichou, Olivier and Tejedor, Vincent and Voituriez, Rapha{\"e}l and Klafter, Joseph},
  journal      = {Nature},
  volume       = {450},
  number       = {7166},
  pages        = {77--80},
  year         = {2007},
  doi          = {10.1038/nature06201}
}

@book{polchinski1998,
place={Cambridge}, 
series={Cambridge Monographs on Mathematical Physics}, 
title={String Theory}, 
publisher={Cambridge University Press}, 
author={Polchinski, Joseph}, 
year={1998}, 
collection={Cambridge Monographs on Mathematical Physics}
}

@article{tenenbaum2000,
  title        = {A global geometric framework for nonlinear dimensionality reduction},
  author       = {Tenenbaum, Joshua B. and de Silva, Vin and Langford, John C.},
  journal      = {Science},
  volume       = {290},
  number       = {5500},
  pages        = {2319--2323},
  year         = {2000},
  doi          = {10.1126/science.290.5500.2319}
}

@article{belkin2003,
  title        = {Laplacian Eigenmaps for Dimensionality Reduction and Data Representation},
  author       = {Belkin, Mikhail and Niyogi, Partha},
  journal      = {Neural Computation},
  volume       = {15},
  number       = {6},
  pages        = {1373--1396},
  year         = {2003},
  doi          = {10.1162/089976603321780317}
}

@article{codling2008,
  title        = {Random walk models in biology},
  author       = {Codling, Edward A. and Plank, Michael J. and Benhamou, Simon},
  journal      = {Journal of The Royal Society Interface},
  volume       = {5},
  number       = {25},
  pages        = {813--834},
  year         = {2008},
  doi          = {10.1098/rsif.2008.0014}
}

@article{vonhippel1989,
  title        = {Facilitated target location in biological systems},
  author       = {von Hippel, Peter H. and Berg, Otto G.},
  journal      = {Journal of Biological Chemistry},
  volume       = {264},
  number       = {2},
  pages        = {675--678},
  year         = {1989},
  doi          = {10.1016/S0021-9258(19)84994-3}
}

@article{peach2022,
  title        = {Relative, local and global dimension in complex networks},
  author       = {Peach, Robert and Arnaudon, Alexis and Barahona, Mauricio},
  journal      = {Nature Communications},
  volume       = {13},
  number       = {1},
  pages        = {3088},
  year         = {2022},
  doi          = {10.1038/s41467-022-30705-w}
}

@article{desy2023,
  title        = {Dimension matters when modeling network communities in hyperbolic spaces},
  author       = {D{\'e}sy, Blaise and Desrosiers, Patrick and Allard, Antoine},
  journal      = {PNAS Nexus},
  volume       = {2},
  number       = {5},
  pages        = {pgad136},
  year         = {2023},
  doi          = {10.1093/pnasnexus/pgad136}
}

@article{vanderkolk2022,
  title        = {An anomalous topological phase transition in spatial random graphs},
  author       = {van der Kolk, Jasper and Serrano, M. {\'A}ngeles and Bogu{\~n}{\'a}, Mari{\'a}n},
  journal      = {Communications Physics},
  volume       = {5},
  number       = {1},
  pages        = {245},
  year         = {2022}
}

@article{van2024random,
  title={Random graphs and real networks with weak geometric coupling},
  author={van der Kolk, Jasper and Serrano, M. {\'A}ngeles and Bogu{\~n}{\'a}, Mari{\'a}n},
  journal={Physical Review Research},
  volume={6},
  number={1},
  pages={013337},
  year={2024},
  publisher={APS}
}

@incollection{farahani2021brief,
  title     = {A Brief Review of Domain Adaptation},
  author    = {Farahani, Ali and Voghoei, Sajad and Rasheed, Khaled and Arabnia, Hamid R.},
  booktitle = {Advances in Data Science and Information Engineering: Proceedings from ICDATA 2020 and IKE 2020},
  publisher = {Springer},
  pages     = {877--894},
  year      = {2021}
}

@article{wang2018deep,
  title   = {Deep Visual Domain Adaptation: A Survey},
  author  = {Wang, Mei and Deng, Weihong},
  journal = {Neurocomputing},
  volume  = {312},
  pages   = {135--153},
  year    = {2018}
}

@inproceedings{tzeng2017adversarial,
  title     = {Adversarial Discriminative Domain Adaptation},
  author    = {Tzeng, Eric and Hoffman, Judy and Saenko, Kate and Darrell, Trevor},
  booktitle = {Proceedings of the IEEE Conference on Computer Vision and Pattern Recognition (CVPR)},
  pages     = {7167--7176},
  year      = {2017}
}

@misc{zenodo,
  author       = {Ferr{\`a} Marc{\'u}s, Aina and Jankowski, Robert and Vila-Mi{\~n}ana, Meritxell and Casacuberta, Carles and Serrano, M. {\'A}ngeles},
  title        = {Chordless cycle filtrations for dimensionality detection in complex networks via topological data analysis},
  year         = {2026},
  month        = apr,
  publisher    = {Zenodo},
  note         = {Dataset. doi: 10.5281/zenodo.19470037. Available at https://doi.org/10.5281/zenodo.19470037}
}

@article{vilaminana2026,
  author  = {Vila-Mi{\~n}ana, Meritxell and Jankowski, Robert and Ferr{\`a} Marc{\'u}s, Aina and Ballester, Rub{\'e}n and Serrano, M. {\'A}ngeles and Casacuberta, Carles},
  title   = {Motif-based filtrations for persistent homology: A framework for graph isomorphism and property prediction},
  journal = {arXiv preprint arXiv:2604.15265},
  volume = {},
  year    = {2026}
}

\onecolumngrid
\clearpage
\includepdf[pages={{},1,2,3,4,5,6,7,8,9,10,11,12,13,14,15,16,17}]{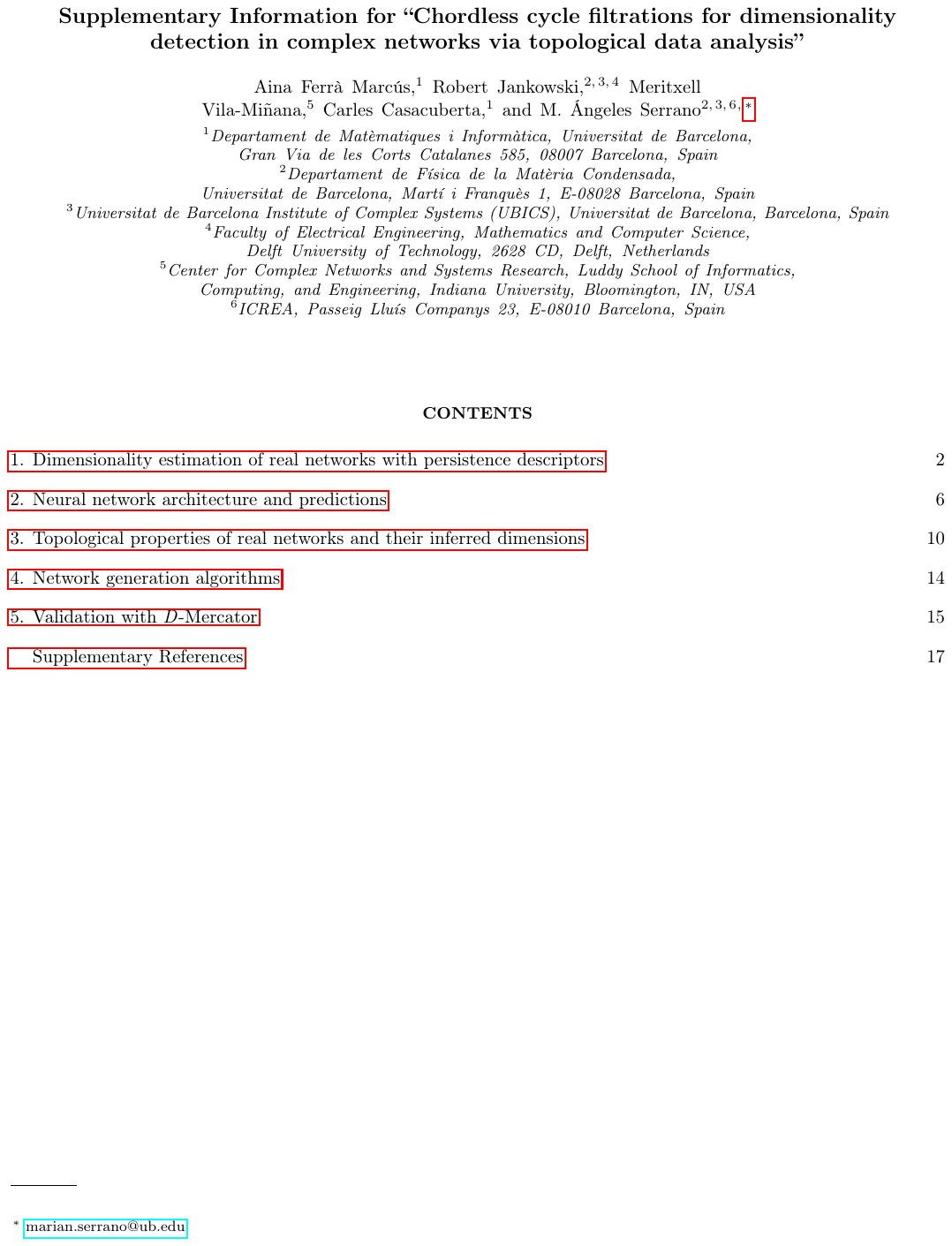}
\end{document}